\documentclass{emulateapj}
\usepackage{graphicx}
\usepackage{mathrsfs}
\usepackage{amsmath}
\usepackage{color}

\shorttitle{Supernova Rates in Known Strong-lens Systems}
\shortauthors{Shu et al. 2018}
 
\begin{document}
 
\title{Prediction of Supernova Rates in Known Galaxy-galaxy Strong-lens Systems}

\author{Yiping Shu$^\dagger$}
\affil{Purple Mountain Observatory, Chinese Academy of Sciences, 2 West Beijing Road, Nanjing 210008, China}
\affil{Institute of Astronomy, University of Cambridge, Madingley Road, Cambridge CB3 0HA, UK ({\tt yiping.shu@pmo.ac.cn})}
\altaffiltext{$^\dagger$}{Royal Society - K. C. Wong International Fellow}

\author{Adam S. Bolton}
\affil{National Optical Astronomy Observatory, 950 N. Cherry Ave., Tucson, AZ 85719, USA ({\tt bolton@noao.edu})}

\author{Shude Mao}
\affil{Physics Department and Tsinghua Centre for Astrophysics, Tsinghua University, Beijing 100084, China}
\affil{National Astronomical Observatories, Chinese Academy of Sciences, 20A Datun Road, Chaoyang District, Beijing 100012, China}
\affil{Jodrell Bank Centre for Astrophysics, School of Physics and Astronomy, The University of Manchester, Oxford Road, Manchester M13 9PL, UK}

\author{Xi Kang}
\affil{Purple Mountain Observatory, Chinese Academy of Sciences, 2 West Beijing Road, Nanjing 210008, China}

\author{Guoliang Li}
\affil{Purple Mountain Observatory, Chinese Academy of Sciences, 2 West Beijing Road, Nanjing 210008, China}

\author{Monika Soraisam}
\affil{National Optical Astronomy Observatory, 950 N. Cherry Ave., Tucson, AZ 85719, USA}

\begin{abstract}
We propose a new strategy of finding strongly-lensed supernovae (SNe) by monitoring 
known galaxy-scale strong-lens systems. Strongly lensed SNe are potentially powerful tools for the study of cosmology, 
galaxy evolution, and stellar populations, but they are extremely 
rare. By targeting known strongly lensed starforming galaxies, our strategy significantly 
boosts the detection efficiency for lensed SNe compared to a blind search. 
As a reference sample, we compile the 128 galaxy-galaxy strong-lens systems from 
the Sloan Lens ACS Survey (SLACS), the SLACS for the Masses Survey, 
and the Baryon Oscillation Spectroscopic Survey Emission-Line Lens Survey. 
Within this sample, we estimate the rates of strongly-lensed Type Ia SN (SNIa) 
and core-collapse SN (CCSN) to be $1.23 \pm 0.12$ and $10.4 \pm 1.1$ events per year, respectively. 
The lensed SN images are expected to be widely separated with a median separation of 2 arcsec. 
Assuming a conservative fiducial lensing magnification 
factor of 5 for the most highly magnified SN image, we forecast that a monitoring program with a single-visit depth of 24.7 mag 
(5$\sigma$ point source, $r$ band) and a cadence of 5 days can detect 0.49 strongly-lensed SNIa event and 
2.1 strongly-lensed CCSN events per year within this sample. 
Our proposed targeted-search strategy is particularly useful for prompt and efficient identifications 
and follow-up observations of strongly-lensed SN candidates. 
It also allows telescopes with small field of views and limited time to efficiently discover 
strongly-lensed SNe with a pencil-beam scanning strategy.

\end{abstract}



\section{Introduction}

Supernovae (SNe) are powerful tools for understanding fundamental physics from 
cosmology to nucleosynthesis. Most prominently for cosmology, the ``standard-candle'' nature of 
Type Ia SN (SNIa) was used to discover the accelerating expansion of the Universe \citep[e.g.,][]{Riess98, Schmidt98, Perlmutter99}. 
By calibrating against distance-independent observables, Type II SNe have also been used 
for cosmological studies \citep[e.g.,][]{Nugent06, Poznanski09, Jones09, Kasen09, Olivares10, 
deJaeger17}, although the dispersion in the distance inferred from Type II SNe ($\sim 10\%$) is 
a factor of $\sim$2 larger than that from SNe Ia \citep[e.g.,][]{Betoule14, Rubin16}. 
Of particular interest for astrophysics, observations of the pre- or early-explosion phases of an SN 
can constrain the properties of its progenitor, and by extension probe the physics of late-stage stellar evolution 
\citep[e.g.,][]{Soderberg08, Nugent11, Li11, Dilday12, Wang13, McCully14, Foley15, 
Ganot16, Tornambe17}. 

The strong gravitational lensing effect furnishes another powerful tool 
for cosmology. From relative delays in the arrival times of multiple images, the Hubble constant can be measured for a given lensing mass model \citep[e.g.,][]{Impey98, Witt00, Treu02, Koopmans03, Suyu10a, 
Suyu13, Suyu14, Wong17, Bonvin17}. This so-called ``time-delay cosmography'' is independent of, 
and complementary to, other cosmological probes such as standard-candle SNIa, baryon acoustic oscillation 
\citep[e.g.,][]{Eisenstein05, Percival10, Anderson14, Zhao17, Wang17, Alam17}, and the fluctuation spectrum of the
cosmic microwave background \citep[e.g.,][]{WMAP9, Planck15}. 
Further observational development of the time-delay cosmography method is
strongly motivated because the the combination of different probes can mitigate the cosmological parameter degeneracies of 
each individual method. Furthermore, 
the lensed source is usually magnified by a factor of ten or more, which makes studies of 
faint objects beyond the detection limit feasible \citep[e.g.,][]{Bolton06b, Quider09, 
Christensen12, Muzzin12, Bussmann13, Stark15, Karman16, Shu16a, Spilker16}. 

The combination of these two tools in the single phenomenon of strongly lensed SNe provides a simultaneous probe of cosmology and of SN physics. All but one time-delay cosmography measurement to date have used the relatively more common phenomenon of strongly lensed 
quasi-stellar objects (QSOs). Compared to lensed QSOs, lensed SNe (and especially lensed SNIa)
are more powerful for cosmological measurements because 1) the optical light curves from which time delays 
are determined are more regular and better understood for SNe \citep[e.g.,][]{Barbon79, 
Doggett85, Kasen09, Sanders15, Morozova17} and 2) the intrinsic luminosity of some SNe can be inferred independently of lensing to constrain the absolute lensing 
magnifications. This extra constraint can be used to break known degeneracies in the lens 
modeling \citep[e.g.,][]{Falco85, Gorenstein88, Saha00, Wucknitz02, Liesenborgs12, Schneider13, 
Schneider14} and lead to a more accurate 
measurement of the Hubble constant, provided the effects from microlensing can be corrected 
\citep[e.g.,][]{Dobler06, Goldstein17b, Yahalomi17, Foxley-Marrable18}.
Furthermore, time delays between lensed images make strong gravitational lensing a natural 
\emph{time machine}. The typical time delays in galaxy-scale strong lenses are on the order of 
1 day to 100 days. One can thus monitor the SN explosion at the predicted location and time 
after detecting the leading lensed image, and catch the trailing images early in the explosive phase. It is in fact possible to watch 
the same explosion multiple times if there are more than two lensed images 
\citep[e.g.,][]{Kelly15, Kelly16}.
A significant sample of strongly-lensed SNe could therefore provide significant improvement in our 
understanding of pre- and early-explosion progenitor properties, especially in combination 
with the lensing magnification effect. 

Despite the great power of strongly-lensed SN systems, there are only three serendipitous 
discoveries to date \citep{Quimby14, Kelly15, Goobar17}. The difficulty 
comes from the fact that a strongly lensed SN system is an $\mathcal{O}$(rare$^3$) event. 
Strong-lensing events (i.e. with multiple lensed images) are rare because they require 
a close alignment between the observer, the lens, and the source. Studies suggest that the 
strong-lens event rate on galaxy scales is on the order of 1 in 1000 \citep[e.g.,][]{SLACSI, 
Marshall09}. At the same time, SN events are rare too. The SN rate in starburst galaxies is 
found to be a few to ten events per century \citep[e.g.,][]{Tammann94, Richmond98, Mannucci03, 
Diehl06, Adams13}. Finally, SNe fade after explosion on rest-frame time scales of 
$\sim 50$ days \citep[e.g.,][]{Richardson02, Li11b}. 
Therefore, observations need to be made in the window before the SNe 
go below the detection limits. 

A large sample of strongly lensed SNe is expected to be discovered by the upcoming 
Large Synoptic Survey Telescope (LSST) project. 
\citet{Oguri10} predict that LSST will discover 46 strongly lensed SNIa and 84 strongly lensed 
core-collapse SN (CCSN, including Type Ib/c and Type II) events with image separations larger than 
0\farcs5 in ten years. 
\citet{Kostrzewa-Rutkowska13} and \citet{Goldstein17} predict furthermore that LSST will detect 
220--1400 strongly-lensed SNIa and $\sim$3800 strongly lensed CCSN events with 
image separations as small as 0\farcs1 in ten years. The numbers of strongly lensed SNe in these latter predictions 
become similar to the prediction of \citet{Oguri10} if limited to events with image separations 
larger than 0\farcs5 and magnifications solely due to brightest images. 
Further limiting to median image separations larger than 2\arcsec, \citet{Goldstein17} predict that 
LSST will detect an average of five (this number is estimated from their Figure 5 as $\sim$20--25\% 
of the total predictions) strongly lensed SNIa events per year. 

In this paper, we propose a new strategy of finding strongly lensed SNe through monitoring of a sample 
of known galaxy-scale strong-lens systems. The advantages of this strategy are three-fold. 
Firstly, any SN occurring at the location of a known
strongly lensed star-forming galaxy has a significantly
higher probability of itself being strongly lensed and multiply imaged, as compared to a SN detected in the field.
Secondly, the multiple SN images detected in this sample will have the same relatively wide image separations as their lensed host galaxies (median of 2\arcsec), 
facilitating intensive early-time observation
of trailing images and precise measurements of
time delays.
Lastly, compared to galaxy clusters behind which previous searches for lensed SNe were carried out 
\citep[e.g.,][]{Goobar09, Postman12, Pan13, Rodney15, Petrushevska16, Petrushevska18}, 
the foreground lens objects of our work are primarily isolated early-type galaxies, 
which have relatively smooth and simple mass profiles that can be more robustly constrained from lensing data. 
This advantage is significant for the application of detected strongly lensed SNe to time-delay cosmography.

This paper is organized as follows. Section~\ref{sect:sample} introduces the compilation 
of strong-lens systems used. Section~\ref{sect:SFR_SNR} describes our procedure for estimating 
the star-formation rates and SN rates in the lensed sources. 
Predictions for detectable SN rates given particular survey depths are presented 
in Section~\ref{sect:detectability}. Discussions and conclusions are given in 
Sections~\ref{sect:discussions} and \ref{sect:conclusions}. 
Throughout the paper, we adopt a fiducial cosmological model with $\rm \Omega_m = 0.308$, 
$\rm \Omega_{\Lambda} = 0.692$ and $H_0 \rm = 67.8\,km\,s^{-1}\,Mpc^{-1}$ 
\citep{Planck15}. 

\section{Strong-lens Sample Compilation}
\label{sect:sample}

To date, almost 300 galaxy-scale strong-lens systems have been discovered  \citep[e.g.,][]{Walsh79, 
Munoz98, Browne03, SLACSV, Faure08, SWELLSI, Brownstein12, More12, 
Inada12, Vieira13, Sonnenfeld13, Stark13, Pawase14, More16a, BELLS_IV, Negrello17, 
Sonnenfeld17, SLACSXIII}. 
In this work, we build a compilation of 128 strong-lens systems from the 
Sloan Lens ACS Survey \citep[SLACS;][63 systems]{SLACSV}, 
the SLACS for the Masses Survey \citep[S4TM;][40 systems]{SLACSXIII}, 
and the Baryon Oscillation Spectroscopy Survey (BOSS) Emission-Line Lens Survey 
\citep[BELLS;][25 systems]{Brownstein12}. 
The strong-lens systems in these three surveys are selected by the  technique introduced 
in \citet{Bolton04}, which ensures that the lensing object is massive early-type galaxy 
while the lensed object is a galaxy with significant star formation as indicated by the 
securely detected [O\textsc{ii}]$\lambda \lambda3727$ emission along with (in many cases) H$\beta$, 
[O\textsc{iii}]4959, [O\textsc{iii}]5007, and H$\alpha$ emission. 
Accurate lens models have been derived from \textsl{Hubble Space Telescope} (\textsl{HST}) 
imaging data for all the systems. The lens and source redshifts are spectroscopically determined. 
The median lens and source redshifts for the SLACS, S4TM, and BELLS lens samples are 
$(z_L=0.20, z_S=0.60)$, $(z_L=0.16, z_S=0.59)$, and $(z_L=0.52, z_S=1.18)$, respectively. 
The median image separation in this compilation is 2\arcsec. 

\section{Star-Formation Rate and Supernova Rates}
\label{sect:SFR_SNR}

The rate of core-collapse SN (CCSN, including Type Ib/c and Type II) in a galaxy, associated with the final collapse of short-lived massive stars, is significantly correlated with the  
star-formation rate (SFR) \citep[e.g.,][]{Madau98, Dahlen99, Hopkins06, Dahlen12, 
Melinder12}. The rate of SNIa depends on other host-galaxy 
properties in addition to SFR \citep[e.g.,][]{Sullivan06, Li11c, Smith12, Graur15}. Following previous work 
\citep[e.g.,][]{Greggio08, Maoz11}, we determine the SNIa rate in our sample by convolving the 
star-formation history (SFH) with the SNIa delay-time distribution, which has been calibrated 
against observations to account for some of the most important host-galaxy property 
dependences (for instance, on host-galaxy mass and age. )


\subsection{Intrinsic [O\textsc{ii}] Flux Determination}

To enable SFR estimation, we first determine the intrinsic integrated [O\textsc{ii}] 
flux of each lensed source, which has been shown to be an empirical 
SFR indicator \citep[e.g.,][]{Gallagher89, Kennicutt92, Kennicutt98, Hogg98, Jansen01, 
Cardiel03, Hopkins03, Kewley04, Moustakas06, Gilbank10, Hayashi13}. We note that a small 
fraction ($\sim 10\%$) of our lensed sources also show 
H$\alpha$ emission in the optical spectroscopic data, which is in principle a better SFR indicator than [O\textsc{ii}] 
\citep[e.g.,][]{Kennicutt93, Kennicutt98, Moustakas06}. 
Nevertheless, for uniformity, we only use [O\textsc{ii}] flux for the SFR estimation for the entire sample. 

The observed integrated [O\textsc{ii}] flux, $F_{\rm [O\textsc{ii}]}$, is measured from 
the SDSS or BOSS spectrum, collected by a fiber centered on the lens galaxy. The lens-galaxy spectrum is 
first subtracted following procedures in \citet{SLACSI}, \citet{Brownstein12}, \citet{Bolton12b}, 
and \citet{BELLSIII}. The [O\textsc{ii}] emission profile is then fitted by a double Gaussian 
plus a first-order polynomial (i.e. a line). The ratio of the central wavelengths of the two Gaussians is 
fixed to $3727.09/3729.88$. 
The parameter optimization is done using the Levenberg-Marquardt algorithm implemented in the 
{\tt MPFIT} package \citep{MPFIT}. The best-fit integrated [O\textsc{ii}] fluxes of 
the lensed sources are provided in Column (7) of Tabel~\ref{tb:tb1}. Note that we choose not 
to report results for 18 systems for which the inferred [O\textsc{ii}] flux error is larger than the 
best-fit [O\textsc{ii}] flux. Calculations of the SFR and SN rates in the rest of the paper 
are based on the remaining 110 strong-lens systems after excluding these 18. 

\begin{figure*}[tbp]
	\includegraphics[width=0.49\textwidth]{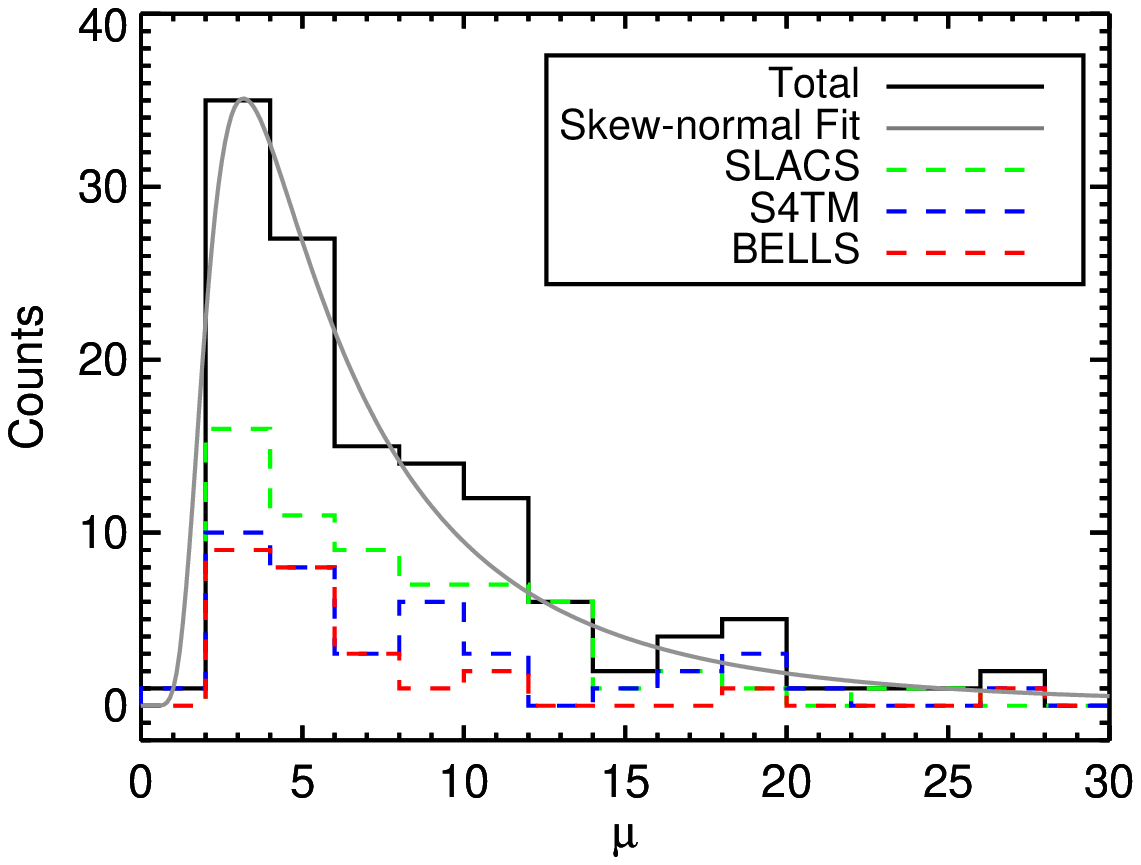}
	\includegraphics[width=0.49\textwidth]{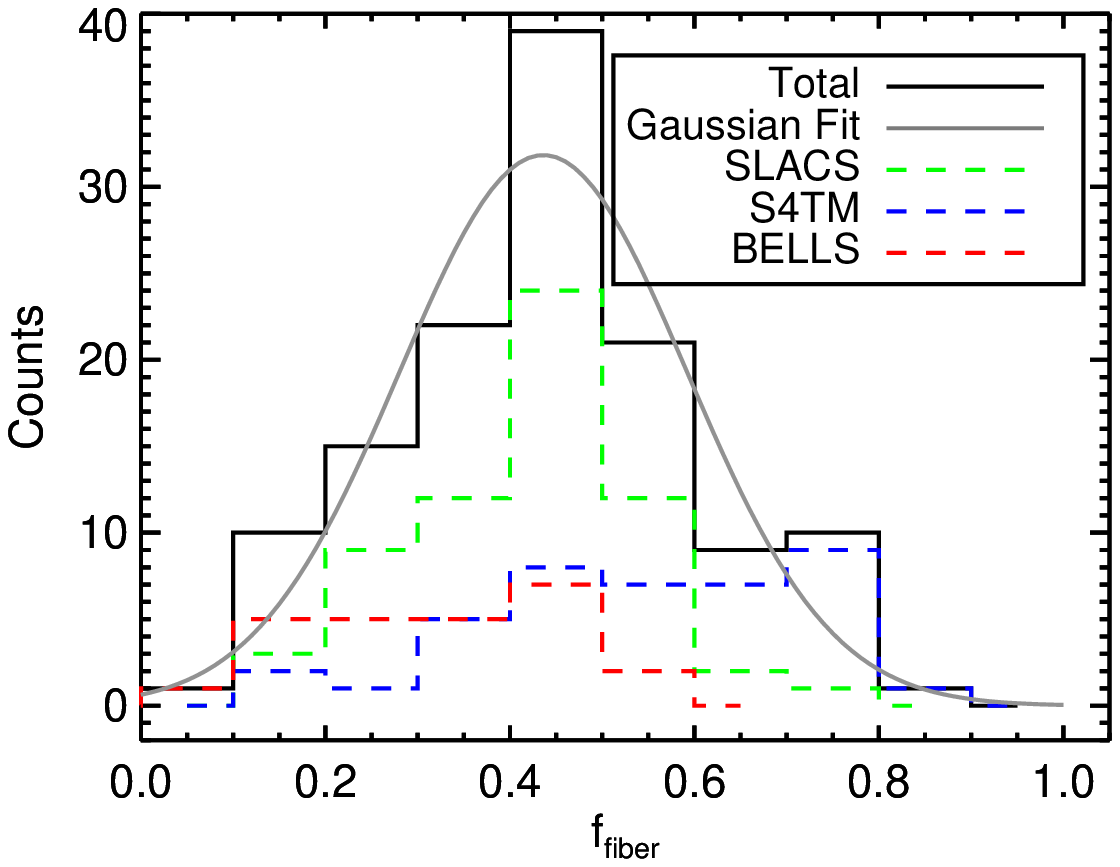}
\caption{\label{fig:dist}
Histograms of the total magnification, $\mu$, and the fraction of flux captured by the fiber, $f_{\rm fiber}$, derived from the best-fit lens models for the full lens compilation (solid black) as well as the three individual lens samples (dashed colors). The gray lines represent the best analytical fits to the full-sample histograms.}
\end{figure*}

To obtain the intrinsic [O\textsc{ii}] integrated flux, we correct for the 
lensing magnification and fiber loss, as inferred from the strong-lens models. For the 
S4TM lensed sources, we use the lens models from \citet{SLACSXIII}. 
We re-model the SLACS strong-lens systems following the same methodology as \citet{SLACSXIII} 
because their lensing magnifications are not publicly available. For consistency, 
we also re-model the BELLS strong-lens systems in the same way. We calculate the total 
magnification, $\mu$, for each lens system. To quantify the uncertainty 
in $\mu$ due to different lens-modeling tools, we compare our values to the published values 
in \citet{Brownstein12} for the 25 BELLS strong-lens systems. The average $\mu$ 
ratio, defined as $\mu_{\rm this\,work}/\mu_{\rm Brownstein2012}$, is $0.94$, 
indicating relatively little bias in the $\mu$ determination. The standard deviation of the 
$\mu$-ratio distribution is $0.16$, which we adopt as the fractional uncertainty in $\mu$. 
The fraction of the lensed-source flux captured by either the $3^{\prime \prime}$-diameter fiber 
(for SLACS and S4TM lenses) or the $2^{\prime \prime}$-diameter fiber (for BELLS lenses), 
denoted as $f_{\rm fiber}$, is determined by convolving the best-fit model of the lensed images 
obtained from the \textsl{HST} imaging data with the ground-based point spread function of the spectroscopic observation. A top-hat kernel matched with the specific fiber size is then applied to the convolved image 
to determine $f_{\rm fiber}$. Columns (5) and (6) of Tabel~\ref{tb:tb1} list $\mu$ and 
$f_{\rm fiber}$ values. The intrinsic source [O\textsc{ii}] integrated flux is then computed as 
\begin{equation}
F_{\rm [O\textsc{ii}]}^{\rm intrinsic} = F_{\rm [O\textsc{ii}]}/f_{\rm fiber}/\mu. 
\end{equation}

Note that all the strong-lens models, and hence the derived $\mu$ and 
$f_{\rm fiber}$ quantities, are based on   
\textsl{HST} F814W wide-band imaging data, which covers roughly 7500\AA--9500\AA\, 
in the observed frame. In the source rest frame, these \textsl{HST} observations probe
continuum emission at 4500\AA--6000\AA\ (for the SLACS and S4TM lensed sources with an 
average redshift of 0.6) and 3400\AA--4300\AA\ (for the BELLS lensed sources with an average 
redshift of 1.2.) Strictly speaking, the derived $\mu$ and $f_{\rm fiber}$ apply to 
the source continuum-emitting regions. We nevertheless adopt these
as the values for the source 
[O\textsc{ii}] emitting regions simply because no other data are available for better determinations 
of $\mu$ and $f_{\rm fiber}$ of [O\textsc{ii}] emission. 
At least for BELLS lensed sources, this approximation should be
reasonably accurate because the 
[O\textsc{ii}] emission is driven by the ultraviolet (UV) radiation 
from massive young stars, which are also primarily responsible for the UV continuum emission around 3400\AA--4300\AA. 

The distributions of $\mu$ and $f_{\rm fiber}$ for the full sample, as well as for each individual lens sample, are shown in Figure~\ref{fig:dist}. 
The $\mu$-histogram of the full sample suggests that the probability density function (PDF) 
of $\log_{10} \mu$ can be well characterized by a skew-normal function as 
\begin{equation}
p(\log_{10} \mu) = \frac{2}{\sigma_{\log_{10} \mu}} \phi(\frac{\log_{10} (\mu/\mu_0)}{\sigma_{\log_{10} \mu}}) \Phi(\alpha (\frac{\log_{10} (\mu/\mu_0)}{\sigma_{\log_{10} \mu}})), 
\end{equation}
in which $\phi(x)$ and $\Phi(x)$ are the PDF and cumulative density function of a Gaussian 
distribution, respectively. The three parameters $\mu_0$, $\sigma_{\log_{10} \mu}$, and $\alpha$
describe the location, scale, and degree of skewness of the $\log_{10} \mu$ PDF (respectively). 
The gray curve in the left panel of Figure~\ref{fig:dist} shows the best skew-normal fit 
to the histogram of the full sample (in black) with 
\begin{align}
\mu_0 &= 5.0 \pm 1.9, \\
\sigma_{\log_{10} \mu} &= 0.91 \pm 0.09, \\
\alpha &= 2.5 \pm 1.0. 
\end{align}
We find that the $\mu$-distributions of the three individual lens samples (as shown by the 
dashed histograms in color) show no significant deviation from that of the full sample.

The $f_{\rm fiber}$ PDF of the full sample can be approximated 
as a Gaussian function with best-fit mean $f_{\rm fiber, 0}$ and standard deviation
$\sigma_{f_{\rm fiber}}$ parameters given by
\begin{align}
f_{\rm fiber, 0} &= 0.44 \pm 0.02, \\
\sigma_{f_{\rm fiber}} &= 0.16 \pm 0.02. 
\end{align}
The distributions of the individual lens samples are similar at small $f_{\rm fiber}$ values, 
but become significantly different at large $f_{\rm fiber}$. 
The BELLS lens sample does not have any system with $f_{\rm fiber} > 60\%$. 
This is primarily related to the fact that the fiber size for BELLS lens systems is more than 
30\% smaller than that of SLACS and S4TM lens systems. On the other hand, over 40\% of the 
S4TM lens systems have $f_{\rm fiber} > 60\%$, which is much larger than the fraction for 
the SLACS lens systems ($<$ 5\%) with the same fiber size. We attribute this 
difference to the fact that the S4TM lens galaxies are, on average, a factor of two less 
massive than the SLACS lens galaxies as found in \citet{SLACSXIII}. With similar lens 
and source redshifts, the less massive S4TM lens galaxies form lensed images with smaller 
angular separations, which are more likely to be covered by the fiber. 

\subsection{Star-formation Rate from [O\textsc{ii}] Luminosity}

Additional calibration is generally needed for an [O\textsc{ii}]-derived SFR because [O\textsc{ii}] 
emission is not directly coupled to the ionizing flux alone, but rather is also sensitive to the oxygen abundance. 
In addition, the intrinsic [O\textsc{ii}] luminosity can not be accurately inferred without 
correction for dust extinction. 
Various studies have attempted to calibrate the [O\textsc{ii}]-derived SFR against 
other observables such as oxygen abundance, [O\textsc{ii}]/H$\alpha$ line ratio, broad-band luminosity, 
stellar mass, and [O\textsc{ii}] equivalent width \citep[e.g.,][]{Hicks02, Kewley04, 
Moustakas06, Argence09, Gilbank10, Mostek12, Hayashi13, Talia15}. 
However, due to the limited data available for the lensed sources, we can not perform a full 
calibration on a system-by-system basis. Instead, we analyze the subsample of our lens 
compilation with detected H$\alpha$ emission, and apply the resulting
calibration to the full sample. 

H$\alpha$ luminosity is one of the most reliable SFR indicators, since it scales directly 
with ionizing flux. A commonly used conversion due to \citet{Kennicutt98} is 
\begin{equation}
\frac{\text{SFR}}{M_{\odot}\text{yr}^{-1}} = 7.9 \times 10^{-42} \frac{L_{\rm H\alpha}}{\text{erg s}^{-1}}, 
\end{equation}
which assumes no dust and a Salpeter initial mass function \citep[IMF,][]{Salpeter55}. 
Here, $L_{\rm H\alpha}$ is the intrinsic H$\alpha$ luminosity. In the presence of dust, 
we can rewrite this relation as
\begin{equation}
\frac{\text{SFR}}{M_{\odot}\text{yr}^{-1}} = 7.9 \times 10^{-42} \frac{10^{0.4 A_{\rm H \alpha}} L_{\rm H\alpha}^{\rm obs}}{\text{erg s}^{-1}}, \nonumber
\end{equation}
in which $L_{\rm H\alpha}^{\rm obs}$ is the observed H$\alpha$ luminosity and 
$A_{\rm H \alpha}$ is the dust extinction at the wavelength of H$\alpha$. We can further transform the 
conversion as 
\begin{equation}
\frac{\text{SFR}}{M_{\odot}\text{yr}^{-1}} = \frac{10^{0.4 A_{\rm H \alpha}}}{F_{\rm [O\textsc{ii}]}/F_{\rm H\alpha}} \frac{L_{\rm [O\textsc{ii}]}^{\rm obs}}{1.27 \times 10^{41} \text{erg s}^{-1}}, \nonumber
\end{equation}
in which $F_{\rm [O\textsc{ii}]}$,  $F_{\rm H\alpha}$, and $L_{\rm [O\textsc{ii}]}^{\rm obs}$ are the observed, dust-extinction uncorrected 
[O\textsc{ii}] flux, H$\alpha$ flux, and [O\textsc{ii}] luminosity, respectively. 
The [O\textsc{ii}] luminosity is simply computed from the [O\textsc{ii}] flux 
$F_{\rm [O\textsc{ii}]}^{\rm intrinsic}$ as 
\begin{equation}
L_{\rm [O\textsc{ii}]} = 4\pi D_{L}^2 (z_S) F_{\rm [O\textsc{ii}]}^{\rm intrinsic},  
\end{equation}
where $D_{L} (z_S)$ is the luminosity distance at the source redshift $z_S$. 

\begin{figure}[tbp]
	\includegraphics[width=0.49\textwidth]{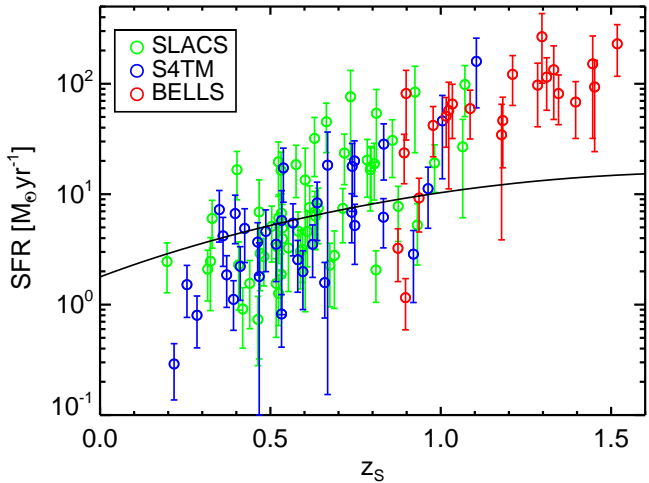}
\caption{\label{fig:SFR_z}
SFR-redshift relation for the SLACS (in green), S4TM (in blue) and BELLS (in red) lensed sources. The black line represents the SFR-redshift relation defined in Equation~(\ref{eq:SFR}) normalized at $z_S=0.9$ where the three lens samples have a substantial overlap. }
\end{figure}

\begin{figure*}[tbp]
	\includegraphics[width=0.49\textwidth]{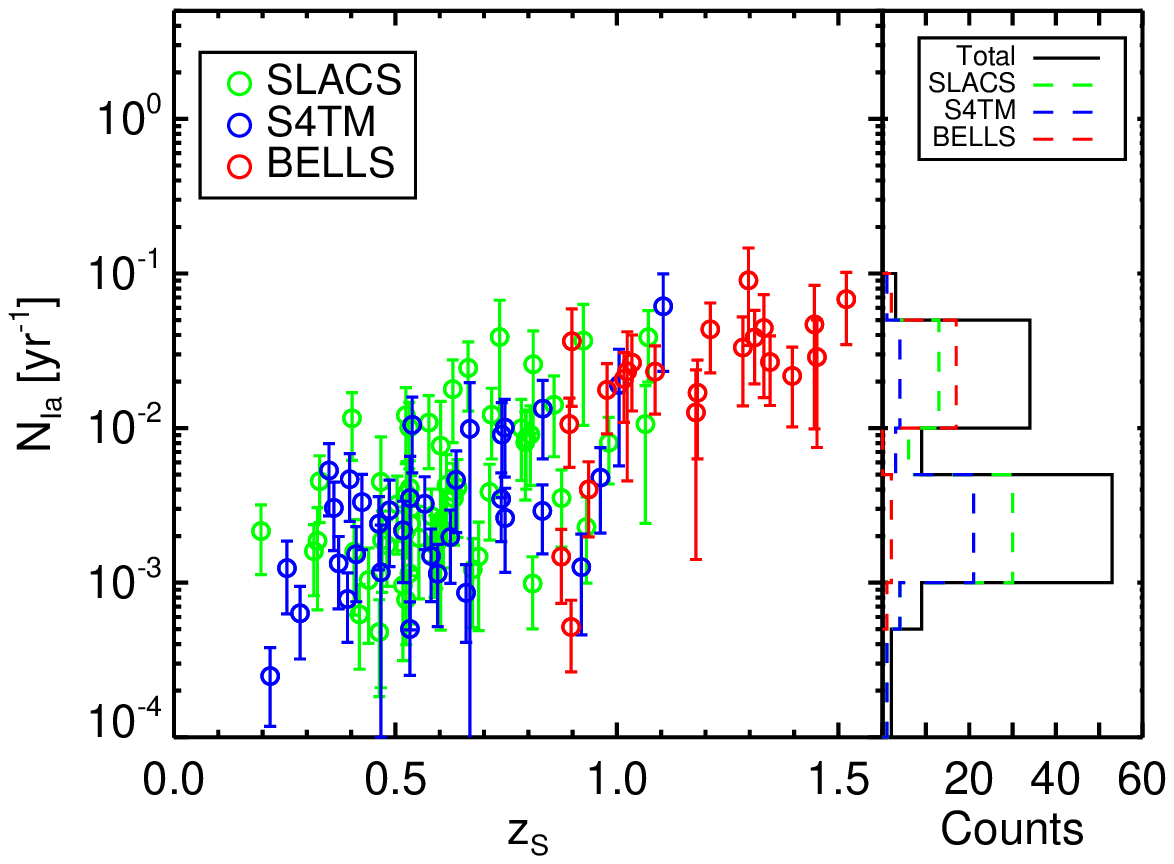}
	\includegraphics[width=0.49\textwidth]{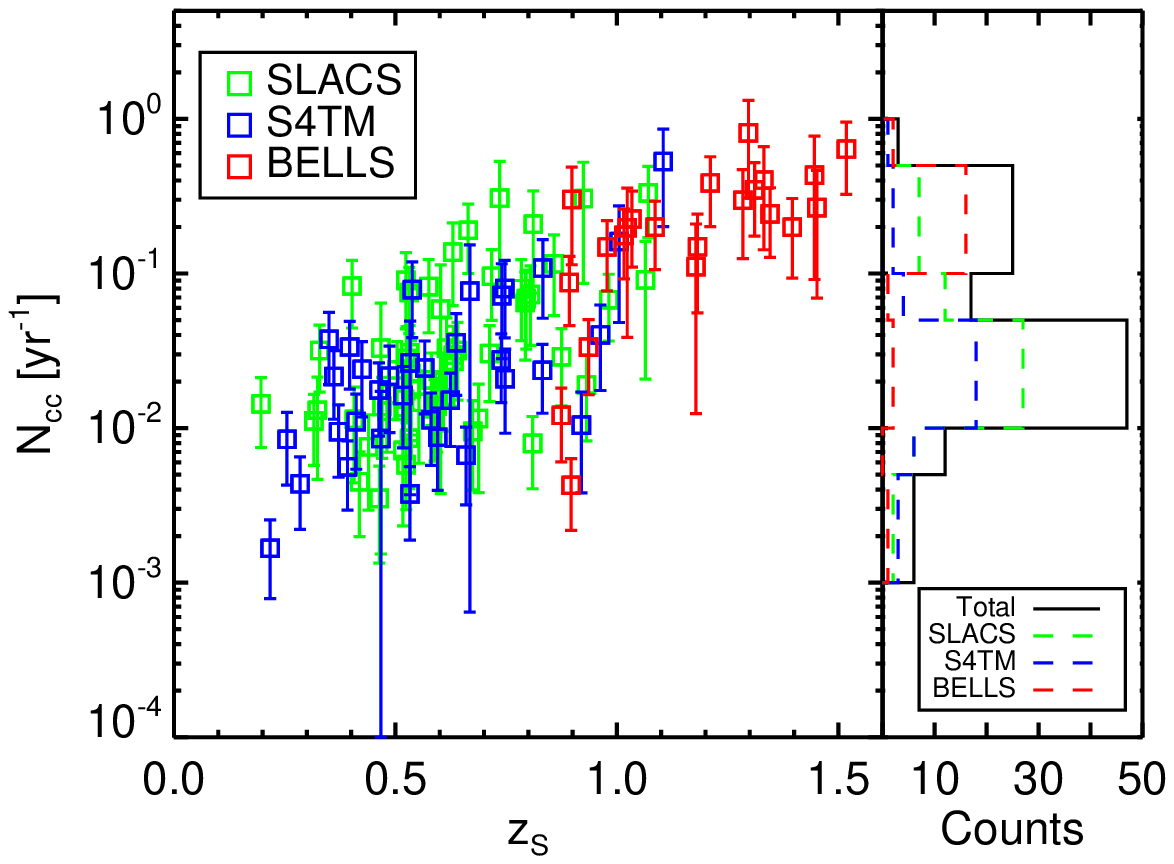}
\caption{\label{fig:n_SN_dist}
Distributions of the expected occurrence rates of SNIa ({\it left}) and CCSN ({\it right}) vs. source redshift for the three lens samples along with the 1-D histograms shown at the side. }
\end{figure*}

We use the 12 strong-lens systems with simultaneously detected H$\alpha$ emissions to 
establish a fiducial [O\textsc{ii}]/H$\alpha$ ratio for the full lens compilation. 
We fit a single Gaussian plus a first-order 
polynomial to the observed H$\alpha$ emission in the residual spectrum to obtain the observed 
H$\alpha$ flux for these 12 lensed sources. We find that the H$\alpha$/[O\textsc{ii}] flux 
ratio is centered at 1.2 with a scatter of 0.5, consistent with the results in 
\citet{Hayashi13}, who studied a sample of $z=1.47$ [O\textsc{ii}] emitters. Furthermore, 
the typical [O\textsc{ii}] luminosity level of the full lensed-source sample is about 
$(4-60) \times 10^{41} $ erg s$^{-1}$, which is also similar to that of the [O\textsc{ii}] 
emitter sample in \citet{Hayashi13}. Based on the H$\alpha$/[O\textsc{ii}] flux ratio, 
\citet{Hayashi13} estimated the dust extinction at H$\alpha$ to be 0.35 mag using the 
empirical relation from \citet{Sobral12}. We thus assume the 
$F_{\rm [O\textsc{ii}]}/F_{\rm H\alpha}$ ratio for the full lens compilation is 
$0.83 \pm 0.35$, and $A_{\rm H \alpha}$ is 0.35 mag. 
The conversion from $L_{\rm [O\textsc{ii}]}$ to SFR now becomes 
\begin{equation}
\frac{\text{SFR}}{M_{\odot}\text{yr}^{-1}} = (1.31 \pm 0.55) \times 10^{-41} L_{\rm [O\textsc{ii}]}. 
\end{equation}
Note that this conversion is obtained assuming a Salpeter IMF, 
while recent studies have suggested that the stellar IMF may vary with 
galaxy properties \citep[e.g.,][]{Treu10a, Auger10, vanDokkum10, Strader11, Cappellari12, 
Sonnenfeld12, Spiniello12, Ferreras13, LaBarbera13, Conroy13, Tortora13, Brewer14, 
Spiniello14, Shu15, Li17}. An extra factor of 0.67 would need to be included to 
convert the reported SFRs to values under the assumption of a Kroupa IMF \citep{Kroupa01}, 
to account for the mass ratio between the two IMFs for the same amount of ionizing radiation 
\citep[e.g.,][]{Brinchmann04, Gilbank10}. 

The estimated intrinsic SFRs for the lensed sources from the above analysis are provided in 
Column (8) of Table~\ref{tb:tb1}. The uncertainty in the SFR takes into account uncertainties 
in the apparent [O\textsc{ii}] flux, total magnification, and conversion factor from 
[O\textsc{ii}] luminosity to SFR. Figure~\ref{fig:SFR_z} shows the estimated SFR against 
redshift of each individual lensed sources from the three lens samples in different colors. 
We find that the SFRs for the lensed sources range from 0.3 
$M_{\odot}$\,yr$^{-1}$ to 267 $M_{\odot}$\,yr$^{-1}$. The median SFRs for the SLACS, 
S4TM, and BELLS lensed sources are 6 $M_{\odot}$\,yr$^{-1}$, 5 $M_{\odot}$\,yr$^{-1}$, 
and 68 $M_{\odot}$\,yr$^{-1}$, respectively. Furthermore, there is 
a strong correlation between SFR and source redshift, which is primarily 
a selection bias, since all the lens systems are selected to have their apparent [O\textsc{ii}] 
fluxes above a certain signal-to-noise ratio (SNR) threshold. One can see from Column (7) in 
Table~\ref{tb:tb1} that the apparent [O\textsc{ii}] flux level is indeed similar across 
the three lens samples. As a result, the intrinsic [O\textsc{ii}] luminosities, and hence 
the derived SFRs, are higher at progressively higher redshifts. 
To further contextualize this, we overplot the cosmic star-formation history trend given by 
\citet[][i.e., Equation~(\ref{eq:SFR}) in this work]{Madau14} in black in 
Figure~\ref{fig:SFR_z}. 
This trend is normalized such that the SFR at $z_S=0.9$, where the three lens samples 
overlap, is equal to the median SFR of the lensed sources within $0.85< z_S < 0.95$. 
It can be seen that the lensed galaxies in our combined sample are more intensely star-forming at higher redshifts, even after taking into 
account the cosmic evolution of the SFR. 

\subsection{Supernova Rates from Star-formation Rate}
\subsubsection{Core-Collapse Supernova Rate}

The rate of CCSNe is proportional to recent SFR
\citep[e.g.,][]{Dahlen99, Oguri10} according to the relation
\begin{equation}
\label{eq:n_cc}
N_{\rm cc} = \frac{1}{1+z_S} k_{\rm cc} \times \text{SFR}, 
\end{equation}
in which $k_{\rm cc}$ is the number of CCSNe per unit mass, and $N_{\rm cc}$ is the 
occurrence rate of CCSNe per year. 
An extra factor of $(1+z_S)^{-1}$ is applied to 
account for the time dilation because the derived SFRs are values in the lensed-source 
rest frame while we are interested in the SN rates in the observed frame (The same factor is 
applied to the SNIa rate calculation below as well). Effectively, $k_{\rm cc}$ is the number of CCSN progenitors per unit mass,
averaged across all recently formed stars. Assuming a particular IMF, $\phi(m)$, $k_{\rm cc}$ can be computed as 
\begin{equation}
k_{\rm cc} = \frac{\int_{M_1}^{M_2} \phi(M) \mathrm{d}M}{\int_{M_{\rm min}}^{M_{\rm max}} M\,\phi(M) \mathrm{d}M},
\end{equation}
in which $M_{\rm min}$ and $M_{\rm max}$ are the lower and upper cut-off masses for the 
considered IMF, and $M_1$--$M_2$ represents the mass range of CCSN progenitors. 
To be consistent with the IMF choice for the SFR estimation, we choose a Salpeter IMF 
with $M_{\rm min} = 0.1 M_{\odot}$ and $M_{\rm max} = 125 M_{\odot}$. The mass range of 
CCSN progenitors is assumed to be 8--50 $M_{\odot}$, since more massive stars likely form 
black holes instead of SNe \citep[e.g.,][]{Tsujimoto97, Hopkins06, Oguri10, 
Dahlen12, Melinder12}. 
Under these assumptions, $k_{\rm cc}$ is computed to be 0.0070 $M_{\odot}^{-1}$. 
Note that \citet{Strolger15} empirically found $k_{\rm cc}$ to be 0.0091 $\pm$ 0.0017 
$M_{\odot}^{-1}$, broadly consistent with the value we adopt here. 

\subsubsection{Type Ia Supernova Rate}
In principle, the occurrence rate of SNIa can be similarly estimated  
\citep[e.g.,][]{Dahlen99, Oguri10} according to
\begin{align}
N_{\rm Ia} &= \eta \, C_{\rm Ia} \frac{\int_{t_{\rm min}}^{t(z_S)} \text{SFR}(t(z_S)-t_{\rm D}) f_{\rm D}(t_{\rm D}; \tau) \mathrm{d}t_{\rm D}}{(1+z_S) \int_{t_{\rm min}}^{t(0)} f_{\rm D}(t_{\rm D}; \tau) \mathrm{d}t_{\rm D}} \\
C_{\rm Ia} &=\frac{\int_{M_1^{\prime}}^{M_2^{\prime}} \phi(M) \mathrm{d}M}{\int_{M_{\rm min}}^{M_{\rm max}} m\,\phi(M) \mathrm{d}M},
\end{align}
in which $M_1^{\prime} = 3 \,M_{\odot}$ and $M_2^{\prime}= 8 \,M_{\odot}$ are the mass limits 
of SNIa progenitors, $\eta$ is the fraction of progenitors that successfully explode as an 
SNIa, and $f_{\rm D}(t_{\rm D}; \tau)$ is the delay-time ($t_{\rm D}$) distribution with a 
characteristic time scale $\tau$ and a minimum delay time $t_{\rm min}$. 
However, this conversion is subject to significant systematics. In particular, $\eta$, $\tau$, 
and $f_{\rm D}(t_{\rm D}; \tau)$ are not well constrained. Usually, $\eta$ is calibrated 
by matching the predicted rate at $z=0$ to the observed SNIa rate in the local 
universe. The typical value for $\eta$ is 4\%--10\% \citep[e.g.,][]{Madau98, Dahlen04, 
Hopkins06, Maoz12}, which depends on the assumed delay-time distribution 
and $\tau$ value. 
To be conservative, we choose $\eta=4\%$. Again choosing a Salpeter IMF, we have 
\begin{equation}
\label{eq:n_Ia}
N_{\rm Ia} = 0.00084 M_{\odot}^{-1} \frac{\int_{t_{\rm min}}^{t(z_S)} \text{SFR}(t(z_S)-t_{\rm D}) f_{\rm D}(t_{\rm D}; \tau) \mathrm{d}t_{\rm D}}{(1+z_S) \int_{t_{\rm min}}^{t(0)} f_{\rm D}(t_{\rm D}; \tau) \mathrm{d}t_{\rm D}}. 
\end{equation}
We assume the SFHs of lensed sources, SFR($t(z)$), follow the best-fit functional form of the observed cosmic 
SFH in \citet{Madau14}, and use the measured SFRs at $z_S$ as normalizations
\begin{equation}
\label{eq:SFR}
\text{SFR}(t(z)) = \text{SFR}(z_S) (\frac{1+z(t)}{1+z_S})^{2.7} \, \frac{[1+(\frac{1+z_S}{2.9})^{5.6}]}{[1+(\frac{1+z(t)}{2.9})^{5.6}]}. 
\end{equation}
As will be shown later, the SNIa rates in this sample are primarily determined by the most recent SFRs, 
and therefore different assumptions on the SFHs will only introduce minor variations in the estimated SNIa rates that are 
negligible compared to the overall uncertainties. 
The delay-time distribution is assumed to be a simple power law with 
no minimum delay time, as suggested in \citet{Maoz12}:
\begin{equation}
f_{\rm D}(t_{\rm D}; \tau) \propto t_{\rm D}^{-1.07}.
\end{equation}

\begin{table*}[htbp]
\begin{center}
\caption{\label{tb:tb1} Useful properties of the compilation of strong lenses.}
\begin{tabular}{c c c c c c c c c c c c c c c}
\hline \hline
Target & Position & $z_L$ & $z_{S}$ & $\mu$ & $f_{\rm fiber}$ & $F_{\rm [O\textsc{ii}]}$ & SFR & $N_{\rm cc}$ & $N_{\rm Ia}$ \\
 & & & & & & & $M_{\odot}$\,yr$^{-1}$ & yr$^{-1}$ & yr$^{-1}$ \\
(1) & (2) & (3) & (4) & (5) & (6) & (7) & (8) & (9) & (10) & \\
\hline
SLACSJ\,0008$-$0004 & 00:08:02.96$-$00:04:08.20 & 0.44 & 1.19 & 5 & 43\% & ... & ... & ... & ... \\ 
SLACSJ\,0029$-$0055 & 00:29:07.77$-$00:55:50.50 & 0.23 & 0.93 & 23 & 61\% & 3.3 $\pm$ 1.1 & 5.3 $\pm$ 3.0 & 0.019 $\pm$ 0.011 & 0.0023 $\pm$ 0.0013 \\ 
SLACSJ\,0037$-$0942 & 00:37:53.22$-$09:42:20.14 & 0.20 & 0.63 & 6 & 34\% & 2.2 $\pm$ 0.5 & 6.3 $\pm$ 3.2 & 0.027 $\pm$ 0.014 & 0.0035 $\pm$ 0.0018 \\ 
SLACSJ\,0044$+$0113$^{*}$ & 00:44:02.90$+$01:13:12.50 & 0.12 & 0.20 & 2 & 53\% & 14.5 $\pm$ 2.4 & 2.5 $\pm$ 1.2 & 0.014 $\pm$ 0.007 & 0.0022 $\pm$ 0.0010 \\ 
SLACSJ\,0109$+$1500 & 01:09:33.74$+$15:00:32.50 & 0.29 & 0.52 & 2 & 55\% & 2.4 $\pm$ 1.0 & 6.1 $\pm$ 3.8 & 0.028 $\pm$ 0.017 & 0.0038 $\pm$ 0.0023 \\ 
SLACSJ\,0157$-$0056 & 01:57:58.94$-$00:56:26.10 & 0.51 & 0.92 & 2 & 28\% & 2.5 $\pm$ 1.4 & 83.9 $\pm$ 60.2 & 0.305 $\pm$ 0.219 & 0.0368 $\pm$ 0.0264 \\ 
SLACSJ\,0216$-$0813 & 02:16:52.53$-$08:13:45.44 & 0.33 & 0.52 & 3 & 26\% & 4.6 $\pm$ 1.1 & 19.6 $\pm$ 10.1 & 0.090 $\pm$ 0.046 & 0.0121 $\pm$ 0.0062 \\ 
SLACSJ\,0252$+$0039 & 02:52:45.21$+$00:39:58.40 & 0.28 & 0.98 & 16 & 53\% & 6.3 $\pm$ 0.6 & 19.2 $\pm$ 8.8 & 0.068 $\pm$ 0.031 & 0.0080 $\pm$ 0.0037 \\ 
SLACSJ\,0330$-$0020 & 03:30:12.14$-$00:20:51.90 & 0.35 & 1.07 & 4 & 47\% & 4.7 $\pm$ 0.9 & 98.3 $\pm$ 47.8 & 0.332 $\pm$ 0.162 & 0.0387 $\pm$ 0.0188 \\ 
SLACSJ\,0405$-$0455 & 04:05:35.42$-$04:55:52.40 & 0.08 & 0.81 & 16 & 73\% & 1.7 $\pm$ 0.3 & 2.1 $\pm$ 1.0 & 0.008 $\pm$ 0.004 & 0.0010 $\pm$ 0.0005 \\ 
SLACSJ\,0728$+$3835 & 07:28:04.95$+$38:35:25.70 & 0.21 & 0.69 & 10 & 52\% & 1.8 $\pm$ 0.9 & 2.8 $\pm$ 1.9 & 0.012 $\pm$ 0.008 & 0.0015 $\pm$ 0.0010 \\ 
SLACSJ\,0737$+$3216 & 07:37:28.44$+$32:16:18.66 & 0.32 & 0.58 & 14 & 52\% & 7.2 $\pm$ 1.5 & 4.6 $\pm$ 2.3 & 0.020 $\pm$ 0.010 & 0.0027 $\pm$ 0.0013 \\ 
SLACSJ\,0822$+$2652 & 08:22:42.32$+$26:52:43.50 & 0.24 & 0.59 & 7 & 46\% & 2.1 $\pm$ 0.5 & 3.2 $\pm$ 1.6 & 0.014 $\pm$ 0.007 & 0.0018 $\pm$ 0.0009 \\ 
SLACSJ\,0841$+$3824 & 08:41:28.81$+$38:24:13.70 & 0.12 & 0.66 & 5 & 30\% & ... & ... & ... & ... \\ 
SLACSJ\,0903$+$4116 & 09:03:15.19$+$41:16:09.10 & 0.43 & 1.06 & 8 & 48\% & 3.1 $\pm$ 1.9 & 26.9 $\pm$ 20.7 & 0.091 $\pm$ 0.070 & 0.0106 $\pm$ 0.0082 \\ 
SLACSJ\,0912$+$0029$^{*}$ & 09:12:05.31$+$00:29:01.18 & 0.16 & 0.32 & 6 & 26\% & 4.4 $\pm$ 2.0 & 2.5 $\pm$ 1.6 & 0.013 $\pm$ 0.008 & 0.0019 $\pm$ 0.0012 \\ 
SLACSJ\,0935$-$0003 & 09:35:43.93$-$00:03:34.80 & 0.35 & 0.47 & 3 & 34\% & 3.2 $\pm$ 2.7 & 6.9 $\pm$ 6.6 & 0.033 $\pm$ 0.031 & 0.0045 $\pm$ 0.0043 \\ 
SLACSJ\,0936$+$0913 & 09:36:00.77$+$09:13:35.80 & 0.19 & 0.59 & 7 & 55\% & 2.4 $\pm$ 0.7 & 3.0 $\pm$ 1.6 & 0.013 $\pm$ 0.007 & 0.0018 $\pm$ 0.0010 \\ 
SLACSJ\,0946$+$1006 & 09:46:56.68$+$10:06:52.80 & 0.22 & 0.61 & 18 & 46\% & 7.0 $\pm$ 1.9 & 4.6 $\pm$ 2.4 & 0.020 $\pm$ 0.011 & 0.0026 $\pm$ 0.0014 \\ 
SLACSJ\,0955$+$0101$^{*}$ & 09:55:19.72$+$01:01:44.40 & 0.11 & 0.32 & 4 & 50\% & 5.2 $\pm$ 0.9 & 2.1 $\pm$ 1.0 & 0.011 $\pm$ 0.005 & 0.0016 $\pm$ 0.0008 \\ 
SLACSJ\,0956$+$5100 & 09:56:29.78$+$51:00:06.39 & 0.24 & 0.47 & 8 & 45\% & 4.6 $\pm$ 1.4 & 2.9 $\pm$ 1.6 & 0.014 $\pm$ 0.007 & 0.0019 $\pm$ 0.0010 \\ 
SLACSJ\,0959$+$0410 & 09:59:44.07$+$04:10:17.00 & 0.13 & 0.54 & 5 & 56\% & 3.9 $\pm$ 1.7 & 4.9 $\pm$ 3.1 & 0.023 $\pm$ 0.014 & 0.0030 $\pm$ 0.0019 \\ 
SLACSJ\,0959$+$4416 & 09:59:00.96$+$44:16:39.40 & 0.24 & 0.53 & 9 & 57\% & 2.6 $\pm$ 0.7 & 1.9 $\pm$ 1.0 & 0.009 $\pm$ 0.005 & 0.0011 $\pm$ 0.0006 \\ 
SLACSJ\,1016$+$3859 & 10:16:22.86$+$38:59:03.30 & 0.17 & 0.44 & 5 & 40\% & 1.8 $\pm$ 0.7 & 1.6 $\pm$ 1.0 & 0.008 $\pm$ 0.005 & 0.0010 $\pm$ 0.0006 \\ 
SLACSJ\,1020$+$1122 & 10:20:26.54$+$11:22:41.10 & 0.28 & 0.55 & 6 & 51\% & 2.6 $\pm$ 1.0 & 3.3 $\pm$ 1.9 & 0.015 $\pm$ 0.009 & 0.0019 $\pm$ 0.0012 \\ 
SLACSJ\,1023$+$4230 & 10:23:32.26$+$42:30:01.80 & 0.19 & 0.70 & 11 & 39\% & ... & ... & ... & ... \\ 
SLACSJ\,1029$+$0420 & 10:29:22.94$+$04:20:01.80 & 0.10 & 0.62 & 4 & 47\% & 2.3 $\pm$ 0.8 & 7.6 $\pm$ 4.3 & 0.033 $\pm$ 0.019 & 0.0043 $\pm$ 0.0024 \\ 
SLACSJ\,1032$+$5322$^{*}$ & 10:32:35.84$+$53:22:34.90 & 0.13 & 0.33 & 3 & 38\% & 9.2 $\pm$ 1.0 & 6.0 $\pm$ 2.8 & 0.032 $\pm$ 0.015 & 0.0045 $\pm$ 0.0021 \\ 
SLACSJ\,1100$+$5329 & 11:00:24.39$+$53:29:13.71 & 0.32 & 0.86 & 13 & 28\% & 6.8 $\pm$ 2.0 & 30.7 $\pm$ 16.5 & 0.116 $\pm$ 0.062 & 0.0142 $\pm$ 0.0076 \\ 
SLACSJ\,1103$+$5322 & 11:03:08.21$+$53:22:28.20 & 0.16 & 0.74 & 3 & 24\% & 5.4 $\pm$ 3.1 & 76.3 $\pm$ 55.5 & 0.308 $\pm$ 0.224 & 0.0388 $\pm$ 0.0283 \\ 
SLACSJ\,1106$+$5228 & 11:06:46.16$+$52:28:37.70 & 0.10 & 0.41 & 25 & 23\% & 8.3 $\pm$ 3.5 & 2.3 $\pm$ 1.4 & 0.011 $\pm$ 0.007 & 0.0016 $\pm$ 0.0010 \\ 
SLACSJ\,1112$+$0826 & 11:12:50.60$+$08:26:10.40 & 0.27 & 0.63 & 4 & 11\% & 2.6 $\pm$ 0.8 & 32.0 $\pm$ 17.6 & 0.137 $\pm$ 0.075 & 0.0179 $\pm$ 0.0098 \\ 
SLACSJ\,1134$+$6027 & 11:34:05.88$+$60:27:13.20 & 0.15 & 0.47 & 2 & 18\% & ... & ... & ... & ... \\ 
SLACSJ\,1142$+$1001 & 11:42:57.35$+$10:01:11.80 & 0.22 & 0.50 & 3 & 42\% & 2.5 $\pm$ 0.7 & 5.1 $\pm$ 2.7 & 0.024 $\pm$ 0.012 & 0.0032 $\pm$ 0.0017 \\ 
SLACSJ\,1143$-$0144$^{*}$ & 11:43:29.64$-$01:44:30.00 & 0.11 & 0.40 & 4 & 18\% & 9.0 $\pm$ 1.1 & 16.6 $\pm$ 7.7 & 0.083 $\pm$ 0.039 & 0.0116 $\pm$ 0.0054 \\ 
SLACSJ\,1153$+$4612 & 11:53:10.79$+$46:12:05.30 & 0.18 & 0.88 & 10 & 59\% & 2.7 $\pm$ 0.7 & 7.8 $\pm$ 4.0 & 0.029 $\pm$ 0.015 & 0.0035 $\pm$ 0.0018 \\ 
SLACSJ\,1204$+$0358 & 12:04:44.07$+$03:58:06.39 & 0.16 & 0.63 & 9 & 49\% & 4.9 $\pm$ 1.4 & 6.9 $\pm$ 3.7 & 0.030 $\pm$ 0.016 & 0.0039 $\pm$ 0.0021 \\ 
SLACSJ\,1205$+$4910 & 12:05:40.44$+$49:10:29.38 & 0.21 & 0.48 & 13 & 46\% & 6.2 $\pm$ 0.6 & 2.7 $\pm$ 1.2 & 0.013 $\pm$ 0.006 & 0.0017 $\pm$ 0.0008 \\ 
SLACSJ\,1213$+$6708 & 12:13:40.58$+$67:08:29.00 & 0.12 & 0.64 & 8 & 32\% & 2.8 $\pm$ 0.8 & 7.4 $\pm$ 3.9 & 0.032 $\pm$ 0.017 & 0.0041 $\pm$ 0.0022 \\ 
SLACSJ\,1218$+$0830 & 12:18:26.70$+$08:30:50.30 & 0.14 & 0.72 & 4 & 30\% & ... & ... & ... & ... \\ 
SLACSJ\,1250$+$0523 & 12:50:28.26$+$05:23:49.07 & 0.23 & 0.80 & 10 & 46\% & 6.2 $\pm$ 1.7 & 17.6 $\pm$ 9.3 & 0.069 $\pm$ 0.036 & 0.0086 $\pm$ 0.0045 \\ 
SLACSJ\,1251$-$0208 & 12:51:35.71$-$02:08:05.17 & 0.22 & 0.78 & 4 & 48\% & 3.4 $\pm$ 1.0 & 20.4 $\pm$ 11.0 & 0.080 $\pm$ 0.043 & 0.0100 $\pm$ 0.0054 \\ 
SLACSJ\,1402$+$6321 & 14:02:28.22$+$63:21:33.34 & 0.20 & 0.48 & 33 & 37\% & ... & ... & ... & ... \\ 
SLACSJ\,1403$+$0006 & 14:03:29.49$+$00:06:41.30 & 0.19 & 0.47 & 3 & 45\% & ... & ... & ... & ... \\ 
SLACSJ\,1416$+$5136 & 14:16:22.34$+$51:36:30.40 & 0.30 & 0.81 & 4 & 22\% & 3.1 $\pm$ 1.4 & 54.0 $\pm$ 34.7 & 0.209 $\pm$ 0.134 & 0.0259 $\pm$ 0.0166 \\ 
SLACSJ\,1420$+$6019 & 14:20:15.85$+$60:19:14.80 & 0.06 & 0.54 & 13 & 49\% & 7.0 $\pm$ 1.5 & 3.9 $\pm$ 2.0 & 0.018 $\pm$ 0.009 & 0.0024 $\pm$ 0.0012 \\ 
SLACSJ\,1430$+$4105 & 14:30:04.10$+$41:05:57.10 & 0.28 & 0.58 & 6 & 35\% & 8.6 $\pm$ 1.8 & 18.5 $\pm$ 9.2 & 0.082 $\pm$ 0.041 & 0.0109 $\pm$ 0.0054 \\ 
SLACSJ\,1432$+$6317 & 14:32:13.34$+$63:17:03.80 & 0.12 & 0.66 & 4 & 38\% & 9.2 $\pm$ 1.4 & 45.3 $\pm$ 21.5 & 0.190 $\pm$ 0.090 & 0.0245 $\pm$ 0.0116 \\ 
SLACSJ\,1436$-$0000 & 14:36:27.54$-$00:00:29.10 & 0.29 & 0.80 & 4 & 43\% & 2.7 $\pm$ 0.8 & 18.9 $\pm$ 10.0 & 0.073 $\pm$ 0.039 & 0.0091 $\pm$ 0.0048 \\ 
SLACSJ\,1443$+$0304 & 14:43:19.62$+$03:04:08.25 & 0.13 & 0.42 & 7 & 41\% & 1.5 $\pm$ 0.5 & 0.9 $\pm$ 0.5 & 0.005 $\pm$ 0.003 & 0.0006 $\pm$ 0.0003 \\ 
SLACSJ\,1451$-$0239 & 14:51:28.19$-$02:39:36.40 & 0.13 & 0.52 & 11 & 45\% & 5.3 $\pm$ 1.1 & 3.3 $\pm$ 1.7 & 0.015 $\pm$ 0.008 & 0.0021 $\pm$ 0.0010 \\ 
SLACSJ\,1525$+$3327 & 15:25:06.70$+$33:27:47.40 & 0.36 & 0.72 & 4 & 32\% & 3.7 $\pm$ 0.7 & 23.6 $\pm$ 11.4 & 0.096 $\pm$ 0.046 & 0.0122 $\pm$ 0.0059 \\ 
SLACSJ\,1531$-$0105 & 15:31:50.08$-$01:05:45.60 & 0.16 & 0.74 & 10 & 28\% & ... & ... & ... & ... \\ 
SLACSJ\,1538$+$5817 & 15:38:12.94$+$58:17:09.69 & 0.14 & 0.53 & 8 & 41\% & 15.0 $\pm$ 1.4 & 16.5 $\pm$ 7.5 & 0.075 $\pm$ 0.035 & 0.0101 $\pm$ 0.0046 \\ 
SLACSJ\,1621$+$3931 & 16:21:32.99$+$39:31:44.60 & 0.24 & 0.60 & 8 & 47\% & 9.7 $\pm$ 8.0 & 13.5 $\pm$ 12.6 & 0.059 $\pm$ 0.055 & 0.0077 $\pm$ 0.0072 \\ 
SLACSJ\,1627$-$0053 & 16:27:46.45$-$00:53:57.56 & 0.21 & 0.52 & 20 & 49\% & 3.7 $\pm$ 0.7 & 1.3 $\pm$ 0.6 & 0.006 $\pm$ 0.003 & 0.0008 $\pm$ 0.0004 \\ 
SLACSJ\,1630$+$4520 & 16:30:28.16$+$45:20:36.28 & 0.25 & 0.79 & 9 & 22\% & 2.7 $\pm$ 1.0 & 16.7 $\pm$ 9.6 & 0.065 $\pm$ 0.038 & 0.0081 $\pm$ 0.0047 \\ 
SLACSJ\,1636$+$4707 & 16:36:02.62$+$47:07:29.57 & 0.23 & 0.67 & 8 & 62\% & 1.5 $\pm$ 0.6 & 2.3 $\pm$ 1.3 & 0.010 $\pm$ 0.006 & 0.0012 $\pm$ 0.0007 \\ 
SLACSJ\,2238$-$0754 & 22:38:40.20$-$07:54:56.05 & 0.14 & 0.71 & 12 & 52\% & 5.3 $\pm$ 1.3 & 7.4 $\pm$ 3.8 & 0.030 $\pm$ 0.016 & 0.0039 $\pm$ 0.0020 \\ 
SLACSJ\,2300$+$0022 & 23:00:53.14$+$00:22:37.95 & 0.23 & 0.46 & 12 & 49\% & 2.0 $\pm$ 0.8 & 0.7 $\pm$ 0.5 & 0.004 $\pm$ 0.002 & 0.0005 $\pm$ 0.0003 \\ 
SLACSJ\,2303$+$1422 & 23:03:21.72$+$14:22:17.91 & 0.16 & 0.52 & 8 & 41\% & 1.6 $\pm$ 0.8 & 1.6 $\pm$ 1.1 & 0.007 $\pm$ 0.005 & 0.0010 $\pm$ 0.0007 \\ 
SLACSJ\,2321$-$0939 & 23:21:20.93$-$09:39:10.27 & 0.08 & 0.53 & 12 & 33\% & 7.6 $\pm$ 1.6 & 6.7 $\pm$ 3.3 & 0.031 $\pm$ 0.015 & 0.0041 $\pm$ 0.0020 \\ 
SLACSJ\,2341$+$0000 & 23:41:11.57$+$00:00:18.70 & 0.19 & 0.81 & 11 & 44\% & ... & ... & ... & ... \\ 
\hline \hline
\end{tabular}
\end{center}
\textsc{      Note.} --- Column 1 is the system name with leading letters indicating the specific survey program. Column 2 is the system position in right ascension and declination. Columns 3 and 4 are the lens and source redshifts inferred from the SDSS/BOSS spectrum. Column 5 provides the inferred total magnification from the best-fit lens model. Columns 6 and 7 are the fraction of source flux and apparent [O\textsc{ii}] flux (in units of 10$^{-16}$ erg\,s$^{-1}$\,cm$^2$) collected by the SDSS/BOSS fiber. Columns 8 is the intrinsic SFR estimated from [O\textsc{ii}] flux. Columns 9 and 10 are the core-collapse SN rate, and Type-Ia SN rate inferred from the SFR. Asterisked are systems with simultaneous H$\alpha$ detections. \\
\end{table*}
\addtocounter{table}{-1}
\begin{table*}[htbp]
\begin{center}
\caption{\textit{Continued}}
\begin{tabular}{c c c c c c c c c c c c c c}
\hline \hline
Target & Position & $z_L$ & $z_{S}$ & $\mu$ & $f_{\rm fiber}$ & $F_{\rm [O\textsc{ii}]}$ & SFR & $N_{\rm cc}$ & $N_{\rm Ia}$ \\
 & & & & & & & $M_{\odot}$\,yr$^{-1}$ & yr$^{-1}$ & yr$^{-1}$ \\
(1) & (2) & (3) & (4) & (5) & (6) & (7) & (8) & (9) & (10) \\
\hline
\phantom{B}S4TMJ\,0143$-$1006 & 01:43:56.58$-$10:06:33.70 & 0.22 & 1.10 & 3 & 33\% & 4.3 $\pm$ 1.8 & 159.5 $\pm$ 99.1 & 0.531 $\pm$ 0.330 & 0.0614 $\pm$ 0.0381 \\ 
\phantom{B}S4TMJ\,0159$-$0006 & 01:59:30.14$-$00:06:12.40 & 0.16 & 0.75 & 6 & 56\% & 1.7 $\pm$ 0.6 & 5.2 $\pm$ 2.9 & 0.021 $\pm$ 0.012 & 0.0026 $\pm$ 0.0015 \\ 
\phantom{B}S4TMJ\,0324$+$0045 & 03:24:15.50$+$00:45:05.50 & 0.32 & 0.92 & 14 & 77\% & 1.5 $\pm$ 0.7 & 2.9 $\pm$ 1.8 & 0.010 $\pm$ 0.007 & 0.0013 $\pm$ 0.0008 \\ 
\phantom{B}S4TMJ\,0324$-$0110 & 03:24:54.50$-$01:10:29.10 & 0.45 & 0.62 & 4 & 79\% & 2.1 $\pm$ 0.4 & 3.5 $\pm$ 1.7 & 0.015 $\pm$ 0.008 & 0.0020 $\pm$ 0.0010 \\ 
\phantom{B}S4TMJ\,0753$+$3416 & 07:53:46.21$+$34:16:33.60 & 0.14 & 0.96 & 27 & 56\% & 6.7 $\pm$ 2.3 & 11.2 $\pm$ 6.3 & 0.040 $\pm$ 0.023 & 0.0048 $\pm$ 0.0027 \\ 
\phantom{B}S4TMJ\,0754$+$1927 & 07:54:28.52$+$19:27:28.10 & 0.15 & 0.74 & 5 & 43\% & 4.1 $\pm$ 1.7 & 17.9 $\pm$ 10.9 & 0.072 $\pm$ 0.044 & 0.0091 $\pm$ 0.0055 \\ 
\phantom{B}S4TMJ\,0757$+$1956 & 07:57:48.99$+$19:56:16.30 & 0.12 & 0.83 & 9 & 32\% & 5.3 $\pm$ 1.4 & 28.4 $\pm$ 15.0 & 0.108 $\pm$ 0.057 & 0.0134 $\pm$ 0.0070 \\ 
\phantom{B}S4TMJ\,0826$+$5630 & 08:26:39.86$+$56:30:35.99 & 0.13 & 1.29 & 85 & 61\% & ... & ... & ... & ... \\ 
\phantom{B}S4TMJ\,0847$+$2348 & 08:47:27.69$+$23:48:19.50 & 0.16 & 0.53 & 17 & 79\% & 3.1 $\pm$ 0.7 & 0.8 $\pm$ 0.4 & 0.004 $\pm$ 0.002 & 0.0005 $\pm$ 0.0003 \\ 
\phantom{B}S4TMJ\,0851$+$0505 & 08:51:41.89$+$05:05:07.00 & 0.13 & 0.64 & 6 & 69\% & 5.9 $\pm$ 1.8 & 8.3 $\pm$ 4.5 & 0.036 $\pm$ 0.019 & 0.0046 $\pm$ 0.0025 \\ 
\phantom{B}S4TMJ\,0920$+$3028 & 09:20:48.28$+$30:28:18.40 & 0.29 & 0.39 & 8 & 48\% & 3.3 $\pm$ 0.5 & 1.1 $\pm$ 0.5 & 0.006 $\pm$ 0.003 & 0.0008 $\pm$ 0.0004 \\ 
\phantom{B}S4TMJ\,0955$+$3014 & 09:55:57.49$+$30:14:50.90 & 0.32 & 0.47 & 4 & 72\% & 2.4 $\pm$ 2.2 & 1.8 $\pm$ 1.8 & 0.009 $\pm$ 0.009 & 0.0012 $\pm$ 0.0012 \\ 
\phantom{B}S4TMJ\,0956$+$5539 & 09:56:54.84$+$55:39:47.30 & 0.20 & 0.85 & 19 & 46\% & ... & ... & ... & ... \\ 
\phantom{B}S4TMJ\,1010$+$3124 & 10:10:26.80$+$31:24:17.60 & 0.17 & 0.42 & 4 & 48\% & 5.0 $\pm$ 1.2 & 4.9 $\pm$ 2.5 & 0.024 $\pm$ 0.012 & 0.0033 $\pm$ 0.0017 \\ 
\phantom{B}S4TMJ\,1031$+$3026 & 10:31:21.01$+$30:26:58.00 & 0.17 & 0.75 & 5 & 69\% & 7.1 $\pm$ 1.9 & 20.0 $\pm$ 10.4 & 0.080 $\pm$ 0.042 & 0.0101 $\pm$ 0.0053 \\ 
\phantom{B}S4TMJ\,1040$+$3626$^{*}$ & 10:40:58.50$+$36:26:28.60 & 0.12 & 0.28 & 3 & 78\% & 3.4 $\pm$ 0.7 & 0.8 $\pm$ 0.4 & 0.004 $\pm$ 0.002 & 0.0006 $\pm$ 0.0003 \\ 
\phantom{B}S4TMJ\,1041$+$0112$^{*}$ & 10:41:22.85$+$01:12:24.20 & 0.10 & 0.22 & 5 & 73\% & 4.2 $\pm$ 1.2 & 0.3 $\pm$ 0.2 & 0.002 $\pm$ 0.001 & 0.0002 $\pm$ 0.0001 \\ 
\phantom{B}S4TMJ\,1048$+$1313 & 10:48:09.40$+$13:13:52.90 & 0.13 & 0.67 & 4 & 45\% & 4.2 $\pm$ 3.7 & 18.3 $\pm$ 18.2 & 0.077 $\pm$ 0.076 & 0.0099 $\pm$ 0.0098 \\ 
\phantom{B}S4TMJ\,1051$+$4439 & 10:51:09.41$+$44:39:08.51 & 0.16 & 0.54 & 3 & 52\% & 8.0 $\pm$ 2.0 & 17.3 $\pm$ 8.8 & 0.079 $\pm$ 0.040 & 0.0105 $\pm$ 0.0054 \\ 
\phantom{B}S4TMJ\,1056$+$4141 & 10:56:57.61$+$41:41:14.60 & 0.13 & 0.83 & 10 & 76\% & 3.2 $\pm$ 0.5 & 6.2 $\pm$ 2.9 & 0.024 $\pm$ 0.011 & 0.0029 $\pm$ 0.0014 \\ 
\phantom{B}S4TMJ\,1101$+$1523 & 11:01:13.13$+$15:23:39.60 & 0.18 & 0.52 & 5 & 48\% & 2.7 $\pm$ 0.8 & 3.5 $\pm$ 1.9 & 0.016 $\pm$ 0.009 & 0.0022 $\pm$ 0.0012 \\ 
\phantom{B}S4TMJ\,1116$+$0729 & 11:16:41.66$+$07:29:45.60 & 0.17 & 0.69 & 4 & 17\% & ... & ... & ... & ... \\ 
\phantom{B}S4TMJ\,1127$+$2312$^{*}$ & 11:27:38.70$+$23:12:44.40 & 0.13 & 0.36 & 8 & 19\% & 5.7 $\pm$ 0.8 & 4.2 $\pm$ 2.0 & 0.022 $\pm$ 0.010 & 0.0030 $\pm$ 0.0014 \\ 
\phantom{B}S4TMJ\,1137$+$1818 & 11:37:28.61$+$18:18:12.40 & 0.12 & 0.46 & 10 & 23\% & 3.5 $\pm$ 0.7 & 3.7 $\pm$ 1.8 & 0.018 $\pm$ 0.009 & 0.0024 $\pm$ 0.0012 \\ 
\phantom{B}S4TMJ\,1142$+$2509 & 11:42:38.23$+$25:09:05.50 & 0.16 & 0.66 & 18 & 83\% & 3.4 $\pm$ 0.9 & 1.6 $\pm$ 0.8 & 0.007 $\pm$ 0.003 & 0.0009 $\pm$ 0.0004 \\ 
\phantom{B}S4TMJ\,1144$+$0436$^{*}$ & 11:44:40.13$+$04:36:50.50 & 0.10 & 0.26 & 5 & 62\% & 10.5 $\pm$ 2.2 & 1.5 $\pm$ 0.8 & 0.008 $\pm$ 0.004 & 0.0012 $\pm$ 0.0006 \\ 
\phantom{B}S4TMJ\,1213$+$2930 & 12:13:03.72$+$29:30:22.40 & 0.09 & 0.60 & 21 & 49\% & 4.2 $\pm$ 1.3 & 2.0 $\pm$ 1.1 & 0.009 $\pm$ 0.005 & 0.0011 $\pm$ 0.0006 \\ 
\phantom{B}S4TMJ\,1301$+$0834 & 13:01:26.88$+$08:34:25.20 & 0.09 & 0.53 & 9 & 64\% & 8.9 $\pm$ 6.5 & 5.8 $\pm$ 5.0 & 0.026 $\pm$ 0.023 & 0.0035 $\pm$ 0.0030 \\ 
\phantom{B}S4TMJ\,1330$+$1750$^{*}$ & 13:30:31.40$+$17:50:40.50 & 0.21 & 0.37 & 4 & 50\% & 2.9 $\pm$ 0.6 & 1.9 $\pm$ 0.9 & 0.009 $\pm$ 0.005 & 0.0013 $\pm$ 0.0007 \\ 
\phantom{B}S4TMJ\,1403$+$3309 & 14:03:09.67$+$33:09:17.80 & 0.06 & 0.77 & 9 & 68\% & ... & ... & ... & ... \\ 
\phantom{B}S4TMJ\,1430$+$6104 & 14:30:34.77$+$61:04:04.80 & 0.17 & 0.65 & 11 & 53\% & ... & ... & ... & ... \\ 
\phantom{B}S4TMJ\,1433$+$2835 & 14:33:51.63$+$28:35:16.40 & 0.09 & 0.41 & 10 & 37\% & 5.3 $\pm$ 1.3 & 2.2 $\pm$ 1.1 & 0.011 $\pm$ 0.006 & 0.0015 $\pm$ 0.0008 \\ 
\phantom{B}S4TMJ\,1541$+$3642 & 15:41:22.27$+$36:42:31.70 & 0.14 & 0.74 & 16 & 53\% & 5.9 $\pm$ 0.9 & 6.9 $\pm$ 3.2 & 0.028 $\pm$ 0.013 & 0.0035 $\pm$ 0.0016 \\ 
\phantom{B}S4TMJ\,1543$+$2202$^{*}$ & 15:43:39.94$+$22:02:23.30 & 0.27 & 0.40 & 3 & 70\% & 10.5 $\pm$ 1.3 & 6.7 $\pm$ 3.1 & 0.033 $\pm$ 0.016 & 0.0047 $\pm$ 0.0022 \\ 
\phantom{B}S4TMJ\,1550$+$2020$^{*}$ & 15:50:10.62$+$20:20:13.50 & 0.14 & 0.35 & 2 & 35\% & 6.3 $\pm$ 1.2 & 7.2 $\pm$ 3.6 & 0.038 $\pm$ 0.018 & 0.0053 $\pm$ 0.0026 \\ 
\phantom{B}S4TMJ\,1553$+$3004 & 15:53:16.14$+$30:04:25.70 & 0.16 & 0.57 & 5 & 75\% & 4.4 $\pm$ 0.9 & 5.5 $\pm$ 2.7 & 0.024 $\pm$ 0.012 & 0.0032 $\pm$ 0.0016 \\ 
\phantom{B}S4TMJ\,1607$+$2147 & 16:07:40.50$+$21:47:11.00 & 0.21 & 0.49 & 2 & 53\% & 1.6 $\pm$ 0.6 & 4.6 $\pm$ 2.6 & 0.022 $\pm$ 0.012 & 0.0029 $\pm$ 0.0017 \\ 
\phantom{B}S4TMJ\,1633$+$1441 & 16:33:44.16$+$14:41:54.90 & 0.13 & 0.58 & 19 & 51\% & 5.5 $\pm$ 1.1 & 2.6 $\pm$ 1.3 & 0.011 $\pm$ 0.006 & 0.0015 $\pm$ 0.0007 \\ 
\phantom{B}S4TMJ\,2309$-$0039 & 23:09:46.36$-$00:39:12.90 & 0.29 & 1.00 & 4 & 37\% & 2.1 $\pm$ 1.1 & 46.1 $\pm$ 32.3 & 0.161 $\pm$ 0.113 & 0.0190 $\pm$ 0.0133 \\ 
\phantom{B}S4TMJ\,2324$+$0105 & 23:24:27.77$+$01:05:58.50 & 0.19 & 0.28 & 8 & 73\% & ... & ... & ... & ... \\ 
BELLSJ\,0151$+$0049 & 01:51:07.37$+$00:49:09.00 & 0.52 & 1.36 & 7 & 43\% & ... & ... & ... & ... \\ 
BELLSJ\,0747$+$4448 & 07:47:34.75$+$44:48:59.30 & 0.44 & 0.90 & 26 & 51\% & 0.8 $\pm$ 0.2 & 1.2 $\pm$ 0.6 & 0.004 $\pm$ 0.002 & 0.0005 $\pm$ 0.0003 \\ 
BELLSJ\,0747$+$5055 & 07:47:24.12$+$50:55:37.50 & 0.44 & 0.90 & 3 & 31\% & 3.9 $\pm$ 1.7 & 81.6 $\pm$ 50.7 & 0.301 $\pm$ 0.187 & 0.0365 $\pm$ 0.0226 \\ 
BELLSJ\,0801$+$4727 & 08:01:05.30$+$47:27:49.61 & 0.48 & 1.52 & 4 & 50\% & 3.5 $\pm$ 0.7 & 229.8 $\pm$ 113.0 & 0.639 $\pm$ 0.314 & 0.0682 $\pm$ 0.0335 \\ 
BELLSJ\,0830$+$5116 & 08:30:49.73$+$51:16:31.80 & 0.53 & 1.33 & 6 & 23\% & 2.3 $\pm$ 1.1 & 134.0 $\pm$ 86.5 & 0.402 $\pm$ 0.260 & 0.0444 $\pm$ 0.0287 \\ 
BELLSJ\,0944$-$0147 & 09:44:27.47$-$01:47:42.40 & 0.54 & 1.18 & 5 & 37\% & 1.3 $\pm$ 1.0 & 34.5 $\pm$ 30.6 & 0.111 $\pm$ 0.098 & 0.0126 $\pm$ 0.0112 \\ 
BELLSJ\,1159$-$0007 & 11:59:44.63$-$00:07:28.20 & 0.58 & 1.35 & 8 & 43\% & 3.7 $\pm$ 0.6 & 81.4 $\pm$ 38.8 & 0.243 $\pm$ 0.116 & 0.0268 $\pm$ 0.0128 \\ 
BELLSJ\,1215$+$0047 & 12:15:04.44$+$00:47:26.00 & 0.64 & 1.30 & 3 & 14\% & 1.8 $\pm$ 0.8 & 266.6 $\pm$ 165.3 & 0.812 $\pm$ 0.504 & 0.0903 $\pm$ 0.0560 \\ 
BELLSJ\,1221$+$3806 & 12:21:51.92$+$38:06:10.50 & 0.53 & 1.28 & 4 & 36\% & 1.8 $\pm$ 0.7 & 97.2 $\pm$ 56.4 & 0.298 $\pm$ 0.173 & 0.0332 $\pm$ 0.0193 \\ 
BELLSJ\,1234$-$0241 & 12:34:27.99$-$02:41:29.60 & 0.49 & 1.02 & 2 & 44\% & 1.8 $\pm$ 0.3 & 50.9 $\pm$ 24.3 & 0.177 $\pm$ 0.084 & 0.0208 $\pm$ 0.0099 \\ 
BELLSJ\,1318$-$0104 & 13:18:29.39$-$01:04:21.60 & 0.66 & 1.40 & 7 & 46\% & 2.2 $\pm$ 0.6 & 68.3 $\pm$ 36.3 & 0.199 $\pm$ 0.106 & 0.0218 $\pm$ 0.0116 \\ 
BELLSJ\,1337$+$3620 & 13:37:51.31$+$36:20:18.10 & 0.56 & 1.18 & 12 & 15\% & 1.6 $\pm$ 0.7 & 46.4 $\pm$ 29.1 & 0.149 $\pm$ 0.093 & 0.0169 $\pm$ 0.0106 \\ 
BELLSJ\,1349$+$3612 & 13:49:10.30$+$36:12:39.70 & 0.44 & 0.89 & 5 & 36\% & 2.1 $\pm$ 0.3 & 23.6 $\pm$ 11.2 & 0.087 $\pm$ 0.041 & 0.0106 $\pm$ 0.0050 \\ 
BELLSJ\,1352$+$3216 & 13:52:18.99$+$32:16:51.80 & 0.46 & 1.03 & 6 & 9\% & 1.0 $\pm$ 0.3 & 65.5 $\pm$ 33.5 & 0.225 $\pm$ 0.115 & 0.0265 $\pm$ 0.0136 \\ 
BELLSJ\,1522$+$2910 & 15:22:09.54$+$29:10:21.90 & 0.56 & 1.31 & 6 & 43\% & 3.8 $\pm$ 0.8 & 114.8 $\pm$ 57.2 & 0.348 $\pm$ 0.173 & 0.0385 $\pm$ 0.0192 \\ 
BELLSJ\,1541$+$1812 & 15:41:18.56$+$18:12:35.10 & 0.56 & 1.11 & 4 & 45\% & ... & ... & ... & ... \\ 
BELLSJ\,1542$+$1629 & 15:42:46.33$+$16:29:51.80 & 0.35 & 1.02 & 3 & 25\% & 1.5 $\pm$ 1.0 & 57.2 $\pm$ 46.0 & 0.198 $\pm$ 0.159 & 0.0233 $\pm$ 0.0187 \\ 
BELLSJ\,1545$+$2748 & 15:45:03.57$+$27:48:05.30 & 0.52 & 1.29 & 4 & 23\% & ... & ... & ... & ... \\ 
BELLSJ\,1601$+$2138 & 16:01:13.27$+$21:38:33.90 & 0.54 & 1.45 & 3 & 26\% & 1.3 $\pm$ 0.8 & 151.4 $\pm$ 119.4 & 0.433 $\pm$ 0.342 & 0.0468 $\pm$ 0.0369 \\ 
BELLSJ\,1611$+$1705 & 16:11:09.80$+$17:05:26.60 & 0.48 & 1.21 & 3 & 34\% & 2.1 $\pm$ 0.3 & 121.7 $\pm$ 58.1 & 0.385 $\pm$ 0.184 & 0.0436 $\pm$ 0.0208 \\ 
BELLSJ\,1631$+$1854 & 16:31:50.33$+$18:54:04.10 & 0.41 & 1.09 & 19 & 12\% & 3.6 $\pm$ 0.5 & 59.5 $\pm$ 27.9 & 0.200 $\pm$ 0.094 & 0.0232 $\pm$ 0.0109 \\ 
BELLSJ\,1637$+$1439 & 16:37:14.58$+$14:39:30.10 & 0.39 & 0.87 & 11 & 41\% & 0.8 $\pm$ 0.2 & 3.2 $\pm$ 1.6 & 0.012 $\pm$ 0.006 & 0.0015 $\pm$ 0.0007 \\ 
BELLSJ\,2122$+$0409 & 21:22:52.04$+$04:09:35.50 & 0.63 & 1.45 & 6 & 19\% & 0.9 $\pm$ 0.5 & 93.4 $\pm$ 69.2 & 0.267 $\pm$ 0.197 & 0.0288 $\pm$ 0.0213 \\ 
BELLSJ\,2125$+$0411 & 21:25:10.67$+$04:11:31.60 & 0.36 & 0.98 & 4 & 18\% & 1.3 $\pm$ 0.2 & 42.0 $\pm$ 20.2 & 0.149 $\pm$ 0.071 & 0.0177 $\pm$ 0.0085 \\ 
BELLSJ\,2303$+$0037 & 23:03:35.17$+$00:37:03.20 & 0.46 & 0.94 & 7 & 28\% & 0.9 $\pm$ 0.2 & 9.2 $\pm$ 4.7 & 0.033 $\pm$ 0.017 & 0.0040 $\pm$ 0.0020 \\ 
\hline \hline
\end{tabular}
\end{center}
\end{table*}

The rates of CCSN and SNIa can now be estimated following Equations~(\ref{eq:n_cc}) 
and (\ref{eq:n_Ia}), and are given in Columns (9) and (10) of Table~\ref{tb:tb1} for each 
individual lensed source. The rate of SNIa, $N_{\rm Ia}$, is 0.0002--0.09 per year in the individual lensed galaxies, 
and the median $N_{\rm Ia}$ for the SLACS, S4TM, and BELLS lensed source are 0.004, 0.003, 
and 0.03 per year, respectively. The rate of CCSN in individual lens galaxies, $N_{\rm cc}$, is 
0.002--0.8 per year, and the median $N_{\rm cc}$ for the SLACS, S4TM, and BELLS lensed sources 
are 0.03, 0.02, and 0.2 per year. The integrated rate of SNIa across the entire lens compilation 
is a factor of $\sim 8$ smaller than that of CCSN. 
Figure~\ref{fig:n_SN_dist} shows the predicted rates of both SN types in the lensed source galaxies, as a function of the 
source redshifts. Because the rates are either directly proportional to (for CCSN) or 
determined to the leading order (for SNIa) by the SFR, $N_{\rm Ia}$ and $N_{\rm cc}$ 
increase with source redshift as a result of the same selection effect discussed above. 
This is further evident in the bimodal distributions of $N_{\rm Ia}$ and $N_{\rm cc}$ seen 
in Figure~\ref{fig:n_SN_dist}, with the two peaks corresponding to the lower-redshift SLACS+S4TM 
systems and the higher-redshift BELLS systems. 
For the full lens compilation with 110 strong-lens systems, we expect in total $10.4 \pm 1.1$ 
strongly lensed CCSN events per year, and $1.23 \pm 0.12$ strongly-lensed SNIa events per year.

\section{Rates of Detectable Strongly-lensed Supernovae}
\label{sect:detectability}

We now proceed to estimate the occurrence rates of strongly-lensed SNe in each lensed source of our 
lens compilation that could be detected given specific depths. 
We assume all the detectable SNe in the lensed galaxies are strongly lensed too. 
We use depths of two dedicated time-domain facilities---the Zwicky 
Transient Facility (ZTF) and LSST---as reference survey depths. 
The single-visit depth is 20.4 mag (5$\sigma$ for point source, $R$ band) for ZTF 
\citep{ZTF}, and 24.7 mag (5$\sigma$ for point source, $r$ band) for LSST \citep{LSST}. 
We thus consider ``Program1'' and ``Program2'' as two nominal monitoring programs of this lens compilation with 
cadences of 5 days and single-visit depths of 20.4 mag and 24.7 mag (both 5$\sigma$ for point source) in the SDSS $r$ band 
\citep{Doi10}, respectively.

The detectable SN rates depend on the SN luminosity function, dust extinction, survey depth, 
and lensing magnification. 
We assume peak absolute magnitudes in Johnson $B$-band, $M_B$, of different SN types are Gaussian 
distributed with means and standard deviations given by \citet[][Table 2]{Richardson14}. 
The relative abundances of CCSN subtypes are taken from the same paper. 
For each lensed source, we convert the $M_B$ distributions of different SN types to distributions 
of apparent magnitude in the $r$ band. 
The conversion between $M_B$ and $r$-band apparent magnitude is done based on SN template spectra 
with SNCosmo \citep{Barbary14}, a Python package that takes K correction and distance modulus into account. 
Specifically, the mean spectrum from \citet{Hsiao07} is used for SNIa, the spectrum of SN2006ep 
\citep{Joubert06} is used for Ib, the spectrum of SN2004fe \citep{Pugh04} is used for Ic, 
the spectrum of SN2007pg \citep{Sako14} is used for IIP and IIL, and the spectrum of 
SN2006ix \citep{Bassett06} is used for IIn. Due to the lack of observed spectra with sufficient 
UV coverage for IIb, we use the Ib spectrum as an approximation. 
As justified in the SFR estimations, we assume for each lensed source a constant extinction of 
$A_{\rm H \alpha}=0.35$ mag and convert $A_{\rm H \alpha}$ to the extinction value in the observed 
$r$ band following a Calzetti extinction law \citep{Calzetti2000}. For the redshift range considered 
in this paper (i.e. 0.2--1.5), the applied extinctions in the observed $r$ band are in the range 0.5--0.8 mag.

\begin{figure}[tbph]
	\includegraphics[width=0.49\textwidth]{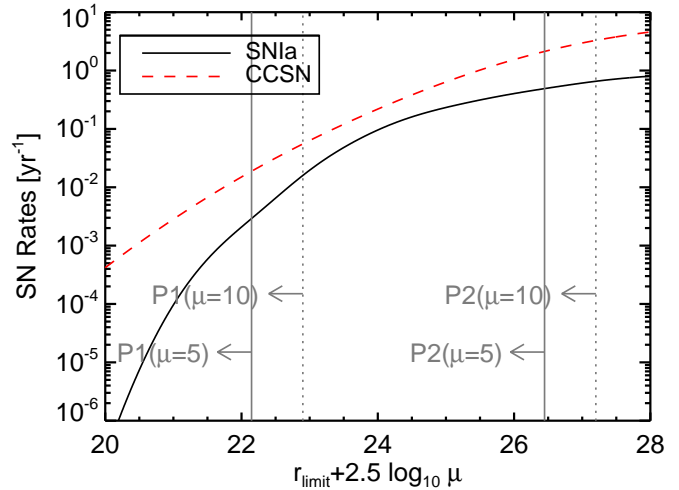}
\caption{\label{fig:SN_vs_r_limit}
Expected occurrence rates of detectable strongly lensed SNIa (black solid line) and CCSN (red dashed line) events as functions of the $r$-band equivalent depth. The vertical gray lines show the equivalent depths of Program1 and Program2 with a fiducial magnification factor of 5 (solid line) and 10 (dotted line).}
\end{figure}

\begin{figure}[tbph]
	\includegraphics[width=0.49\textwidth]{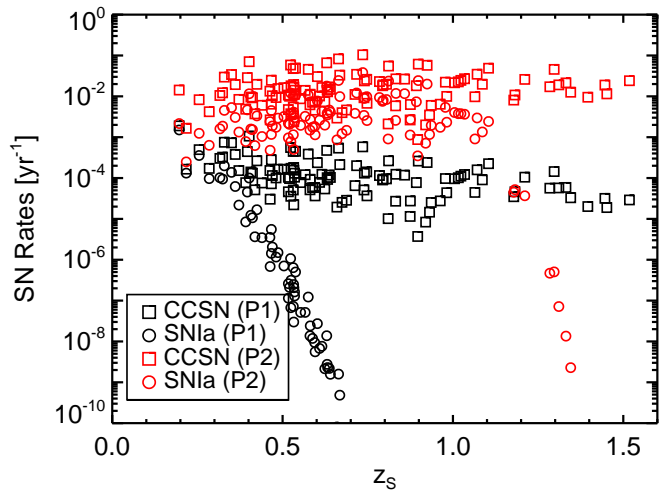}
\caption{\label{fig:SN_vs_z_source}
Expected occurrence rates of strongly-lensed SNIa (circles) and CCSN (squares) events in individual lensed sources that can be detected in Program1 (black symbols) or Program2 (red symbols) as functions of the source redshift assuming a fiducial magnification factor of 5.}
\end{figure}

The lensing magnification essentially extends the instrumental detection limit by a factor 
of 2.5 $\log_{10} \mu$. Although the total magnifications for the extended source continuum 
regions have been calculated from the \textsl{HST} data, 
the expected magnifications of lensed SN images are very uncertain because an SN is 
essentially a point source, the magnification of which is extremely sensitive to its 
actual position in the source plane. It is also possible that the SN explosion happens away 
from the bright continuum region of its host galaxy, especially for SNIa. For the sake of simplicity, 
we adopt for all the sources a fiducial magnification factor of 5 for the most 
magnified SN image, which will extend the detection limit by roughly 1.75 magnitude. 
We note that this is a conservative assumption as the total magnifications of the 
three discovered strongly-lensed SNe (all with four lensed images) are $\sim 30-100$ 
\citep{Quimby14, Kelly15, Goobar17}. 
We choose not to consider magnitude fluctuations induced to individual lens systems by 
microlensing \citep[e.g.,][]{Dobler06} as it would be beyond the scope of this paper. 

In this work, we define a lensed SN as \emph{detected} if its extinction-corrected peak 
$r$-band apparent magnitude is at least 0.7 mag brighter than the program's effective depth, 
defined as the single-visit depth plus 2.5 $\log_{10} \mu$. The detected fractions of all SNe
are derived from the $r$-band apparent-magnitude distributions for every single-visit depth. 
The detectable SN rates in each lensed source are determined by multiplying the total SN rates 
($N_{\rm Ia}$ and $N_{\rm cc}$) by these detected fractions. 
The requirement to be 0.7 mag brighter than the 
effective survey depth, used in \citet{Oguri10}, allows detections of SN 
$\sim$10 rest-frame days before peak \citep[e.g.,][]{Firth15, Hoeflich17, Taddia18, He18}. 
Given a cadence of 5 days, which corresponds to $\sim$ 3 rest-frame days in our sources, this requirement 
ensures a reasonable sampling of the SN light curve, especially the early, pre-maximum color curve that 
is suggested to be insensitive to the microlensing effect and therefore crucial for accurate time-delay 
measurements \citep{Goldstein17b}.

Figure~\ref{fig:SN_vs_r_limit} shows the expected occurrence rates of detectable strongly lensed 
SNIa (black solid line) and CCSN (red dashed line) events as functions of the $r$-band equivalent 
depth. Assuming a fiducial magnification factor of 5, it suggests that in this lens compilation, 
Program1 could detect approximately 0.003 strongly lensed SNIa event and 0.019 strongly-lensed CCSN 
event per year while Program2 could detect 0.49 strongly-lensed SNIa event and 
2.1 strongly-lensed CCSN events per year. 
In terms of the yields in individual lens samples, the expected occurrence rates of detectable 
strongly-lensed SNIa and CCSN in Program2 are 0.32 and 1.2 event per year 
for the SLACS sample, 0.12 and 0.5 event per year for the S4TM sample, and 0.05 and 0.4 
event per year for the BELLS sample. We also plot the 
expectations with a fiducial magnification factor of 10, which are shown by the vertical, 
gray dotted lines. The numbers increase to 0.016 SNIa and 0.055 CCSN event per year in Program1, 
and 0.65 SNIa and 3.3 CCSN events per year in Program2. 

We further examine the expected occurrence rates of detectable strongly lensed SNIa and CCSN 
in each individual lensed source in Figure~\ref{fig:SN_vs_z_source}. 
The black symbols show the rates of detectable SNIa (circle) and 
CCSN (square) as functions of the SN (i.e. source) redshift for Program1 assuming a fiducial 
magnification factor of 5, while the red symbols are the same distributions but for Program2. 
For Program1, the detectabilities of both SN types drop almost monotonically 
against the SN redshift, even though the overall SN rates are higher at higher redshifts. 
The reason is that the decrease in the fractions of detectable SNe overwhelms the increase 
in the overall SN rates at progressively higher redshifts given Program1's shallow detection depth. 
For Program2 with a deeper detection depth, the detectabilities mildly increase with redshift until 
about $z=1$ beyond which the detectability of SNIa starts dropping. 

Our results suggest that it is better to preferentially focus on lower source-redshift strong-lens systems 
in a survey with a limiting magnitude similar to that of Program1. A survey with a depth 
comparable to Program2 is generally sensitive to the entire source-redshift range considered here. 
To further quantify this, we check the slopes of the detectable SN rate-SN redshift relations of both types 
at different detection depths. We find that the slopes become positive when the detection depth reaches 
about 24.5 mag or brighter. This implies that the actual strategy of monitoring for 
strongly lensed SNe in our lens compilation should be adjusted according to the survey depth. 

\section{Discussion}
\label{sect:discussions}

The biggest systematic uncertainty in our SN-rate estimation is in the conversion from 
[O\textsc{ii}] flux to SFR. 
Due to the lack of data for full calibrations on a system-by-system basis, we decided 
to use a subsample with simultaneous detections of H$\alpha$ emissions from the lensed 
sources and apply the resulting calibrations to the full sample. 
We note that the obtained SFR-[O\textsc{ii}] flux relation after calibrations is almost 
identical to the classical relation given by \citet{Kennicutt98}. 
In principle, the uncertainty of this 
calibration, which is found to be 50\% based on current data, can be improved by carrying 
out additional photometric and spectroscopic observations of the lensed sources. 

When deriving SNIa rates, the SFH of each individual lensed sources 
is assumed to follow the cosmic star-formation density evolution trend given in \citet{Madau14} up to an overall
normalization factor. 
In practice, however, the SFH for each individual galaxy will not follow 
this cosmic evolution trend exactly, since it only represents a global average across all galaxies. We therefore investigate the uncertainty introduced to the inferred SNIa rate 
by the choice of SFH by computing the SNIa rate under the 
assumption of a constant SFR until the point of observation for each lensed sources. 
Compared to the SNIa rates reported above, the alternative SNIa rates vary from $-$20\% to $+$3\%, 
with a median relative difference of $-$10\% across all lensed sources. This level of variation is 
much smaller than the systematic uncertainty in the SNIa rate of $\sim$50\%. 
The integrated change in the total SNIa rate of the entire lens compilation is as relatively small $-4\%$. 
We find that the SNIa rate is primarily determined by the most recent SFH 
because of the rapid decline in the delay time function. 
In both SFH scenarios considered, the cumulative SNIa rates have 
already reached 50\% and 90\% within the recent 0.03 and 2 Gyr. 
In fact, the SNIa rate scales almost linearly with the observed SFR with a slope of 
$4.5 \times 10^{-4}$ SNe yr$^{-1}$ ($M_{\odot}$\,yr$^{-1}$)$^{-1}$, consistent 
with previous findings \citep[e.g.,][]{Sullivan06, Smith12, Graur15}. 
Overall, this test suggests that different choices of the SFH would only 
introduce a small uncertainty in the SNIa rate, especially in the aggregate across the entire lens compilation. 

Compared to a blind search, our proposed targeted search is significantly more efficient in 
finding strongly lensed SN events, especially ones with wide image separations. 
For example, 130 strongly lensed SN events 
with image separations larger than 0\farcs5 will be detected by LSST after a ten-year blind search of 
$\sim$20,000 square degrees' sky \citep{Oguri10}. 
Considering the $\sim10$ square degree field of view of LSST, at least 2000 frames need to be 
monitored to yield these SN detections. The detection efficiency, defined as the yield of 
strongly lensed SN events per exposure time, of this blind search is thus $\sim0.006$ event 
per year per frame. 
Our proposed Program2 strategy, with the same depth as LSST but targeted at known strong lenses, has a detection efficiency of 
$\sim0.025$ event per year per frame, and the expected median image separation is 2\arcsec. 
Furthermore, none of the previous predictions for the number of strongly lensed SNe in LSST considered 
the effect of dust extinction, which is known to significantly reduce the number of detections. 
(Our detectable SN rates and detection efficiencies would increase by a factor of 
$\sim$1.5 at LSST depth if the applied dust extinctions were removed. )
Thousands of new strong-lens systems will be discovered in the near future from dedicated surveys 
\citep[e.g.,][]{Marshall09, Oguri10, McKean15, Collett15, Diehl17}. 
The application of an efficient strategy such as ours to the monitoring of these new lenses will significantly 
increase the yield of strongly lensed SNe. 

A sample of strongly lensed SNe, especially SNIa, will be a powerful resource for cosmological 
studies. Time-delay cosmography with lensed QSOs is 
limited by the microlensing effect, which causes uncorrelated variability in the fluxes of different lensed 
images in an unpredictable manner and therefore complicates the time-delay measurement \citep[e.g.,][]{Blandford92, Wyithe02, Dobler06, Morgan12, Tewes13, 
More17, Yahalomi17, Tie18}. 
Compared to QSOs, for which light curves are typically stochastic with variability timescales extending to years, SNe have single-peak, more homogeneous and standardizable light curves that develop on timescales of weeks. 
It is therefore much easier to determine time delays for lensed SNe than lensed QSOs in the presence of the microlensing effect. Furthermore, 
\citet{Goldstein17b} recently found that accurate time-delay measurements can be obtained from 
early-time color curves of strongly-lensed SNIa, which are suggested to be almost immune to 
the microlensing effect. Multiply lensed SN images increase the effective number of SN events for
observational study. In addition, with the aid of lensing 
magnification, more SNe at higher redshifts with better measured light curves can be obtained, 
which can enable more extensive study of the expansion history of the universe at earlier times. 

The standard-candle nature of SNe also provides extra constraining power on 
the mass distribution of the lens galaxy, which holds important clues to galaxy formation and 
evolution \citep[e.g.,][]{Gnedin04, Nipoti04, Gustafsson06, Romano-Diaz08, Governato10, 
Duffy10, Abadi10, Martizzi12, Bolton12, Dubois13, Velliscig14, Shu15}. More accurate 
lens-galaxy mass distributions could be obtained when the mass-sheet and source-position 
transformation degeneracies are broken with known lensing magnifications. 
However, absolute lensing magnifications are not direct observables but rather are degenerate with 
the intrinsic luminosities of the lensed source, which in most cases are unknown. 
Strongly-lensed SNe with intrinsic luminosities that can be derived from lensing-independent 
observables are an important exception, and hence can provide better constraints on the 
lens-galaxy mass distributions \citep[e.g.,][]{Witt95, Oguri03, Bolton03, Mortsell05}. 

Multiply imaged SNe can furthermore revolutionize the observational study of SN early and
pre-explosion phases, which are driven by simple, well understood physics and thus provide 
crucial constraints on the progenitor properties and SN explosion mechanisms. 
For instance, \citet{Pritchard14} and \citet{Ganot16} highlight the importance of early 
UV emissions in CCSNe in determining the physical properties of the progenitors. 
\citet{Tornambe17} suggest that X-ray features in advance of SNIa explosions are 
determined by the rotation of the progenitors, which can be used to distinguish between 
different progenitor scenarios. 
Such types of observations are currently limited to a handful of unlensed SNe 
for which archival data or prompt follow-up observations are available 
\citep[e.g.,][]{Gal-Yam11, Nugent11, Li11, Dilday12, Cao13, Gal-Yam14, Kelly14, McCully14}. 
The delays in the arrival times of multiple SN images, typically on scales from 1 day to 
100 days, essentially replay the SN explosions multiple times. The approximate explosion 
times and exact locations can be predicted from the lens model after the arrival of
the leading SN image. Intensive real-time monitoring of the SN site before its reappearance can 
be scheduled, which will greatly expand the possibilities for early and pre-explosion SN observations. 

\section{Conclusions}
\label{sect:conclusions}

In this paper, we propose a highly efficient strategy of finding strongly lensed SNe by 
monitoring known galaxy-scale strong-lens systems. 
We assemble a compilation of 128 galaxy-galaxy strong-lens systems from 
the SLACS Survey, the S4TM Survey, and the BELLS Survey. All the lensed sources in this 
lens compilation are galaxies at redshifts of 0.2--1.5 with significant 
recent star formation. The median lensed image separation in the sample is 2\arcsec.

The main findings based on this particular lens compilation are as follows. 
\begin{enumerate}
	\item Assuming a Salpeter IMF, the intrinsic SFRs of the lensed sources determined from 
	the observed [O\textsc{ii}] fluxes range from 0.3 $M_{\odot}$\,yr$^{-1}$ to 267 
	$M_{\odot}$\,yr$^{-1}$. The lensing magnification and fiber loss are corrected based on 
	lens models from the \textsl{HST} imaging data. The median SFRs for the SLACS, S4TM, 
	and BELLS lensed sources are 6 $M_{\odot}$\,yr$^{-1}$, 5 $M_{\odot}$\,yr$^{-1}$, 
	and 68 $M_{\odot}$\,yr$^{-1}$, respectively. We find the SFRs become higher at higher 
	redshifts, which is primarily determined by the selection effect that sources at higher 
	redshifts are biased to have higher intrinsic [O\textsc{ii}] flux given a fixed observer-frame
    detection limit. 
	\item The expected overall occurrence rates of SNIa and CCSN in the lensed sources are 
	estimated either directly from the SFRs (for CCSN) or by convolving fiducial SFHs with a 
	SNIa delay-time distribution. For individual lensed sources, the SNIa rates are in the range 
	0.0002--0.09 events per year, and the CCSN rates are in the range 0.002--0.8 events per year. 
	On average, the SNIa rate is a factor of $\sim 8$ lower than the CCSN rate. 
	For the full lens compilation, we predict $1.23 \pm 0.12$ strongly lensed SNIa events per year, 
	and $10.4 \pm 1.1$ strongly lensed CCSN events per year. Mainly due to the same selection 
	effect, the SN rates are higher at progressively higher redshifts. 
	\item We quantify the expected occurrence rates of detectable strongly lensed SNIa and CCSN events 
	in two nominal monitoring programs with different depths. 
	We assume a conservative fiducial magnification factor of 5 for the most magnified 
	SN image is assumed. Within this lens compilation, our Program1 is expected to detect 
	approximately 0.003 strongly-lensed SNIa event and 0.019 strongly-lensed 
	CCSN event per year, while our Program2 can detect 0.49 strongly-lensed SNIa event and 
	2.1 strongly-lensed CCSN events per year. This detection efficiency is significantly higher 
	than that of a blind search.
	\item For this lens compilation, 
	it is more efficient to target either higher-redshift or lower-redshift strong-lens systems 
	depending on whether the single-visit $r$-band depth is brighter than 24.5 mag or not.
\end{enumerate}

In closing, we emphasize that the fundamental principle of our proposed strategy is to focus on 
small targeted sky areas---where known strong-lens systems are located---as being more favorable sites for 
finding strongly-lensed SNe. In the era that hundreds of new SNe (including strongly-lensed SNe) 
are expected to be discovered every night by high-cadence, all-sky surveys such as ZTF and LSST, 
our proposed strategy simply helps in prompt and efficient identifications of the strong-lensing 
nature of the detected SNe. 
Meanwhile, it allows telescopes with small fields of views and limited observing time to efficiently 
discover strongly lensed SNe with a pencil-beam monitoring strategy. As we show in the paper, 
such tageted-search programs with depths comparable to that of LSST can be productive today, which is 
particularly worthwhile considering that LSST is scheduled to start operation in 2022 \citep{LSST17}. 

\acknowledgments

We thank the anonymous referee for comments that substantially improve the manuscript. 
We are grateful for the helpful discussions with Matt Auger, Peter Brown, Kyle Dawson, 
Mark Dickinson, Thomas Matheson, Gautham Narayan, Justin Roberts-Pierel, Steven Rodney, 
Hanna Sai, Lou Strolger, and Danfeng Xiang. 
Y.S. has been supported by the National Natural Science Foundation of China 
(NSFC; No. 11603032 and 11333008), the 973 program (No. 2015CB857003), and 
the Royal Society – K.C. Wong International Fellowship (NF170995). 
S.M. acknowledges the support from the NSFC (Grant No. 11333003, 11390372). 
G.L. is supported by the NSFC (No.11673065, 11273061).

\bibliographystyle{apj}

\begin{thebibliography}{}
\expandafter\ifx\csname natexlab\endcsname\relax\def\natexlab#1{#1}\fi

\bibitem[{{Abadi} {et~al.}(2010){Abadi}, {Navarro}, {Fardal}, {Babul}, \&
  {Steinmetz}}]{Abadi10}
{Abadi}, M.~G., {Navarro}, J.~F., {Fardal}, M., {Babul}, A., \& {Steinmetz}, M.
  2010, \mnras, 407, 435

\bibitem[{{Adams} {et~al.}(2013){Adams}, {Kochanek}, {Beacom}, {Vagins}, \&
  {Stanek}}]{Adams13}
{Adams}, S.~M., {Kochanek}, C.~S., {Beacom}, J.~F., {Vagins}, M.~R., \&
  {Stanek}, K.~Z. 2013, \apj, 778, 164

\bibitem[{{Alam} {et~al.}(2017){Alam}, {Ata}, {Bailey}, {Beutler}, {Bizyaev},
  {Blazek}, {Bolton}, {Brownstein}, {Burden}, {Chuang}, {Comparat}, {Cuesta},
  {Dawson}, {Eisenstein}, {Escoffier}, {Gil-Mar{\'{\i}}n}, {Grieb}, {Hand},
  {Ho}, {Kinemuchi}, {Kirkby}, {Kitaura}, {Malanushenko}, {Malanushenko},
  {Maraston}, {McBride}, {Nichol}, {Olmstead}, {Oravetz}, {Padmanabhan},
  {Palanque-Delabrouille}, {Pan}, {Pellejero-Ibanez}, {Percival}, {Petitjean},
  {Prada}, {Price-Whelan}, {Reid}, {Rodr{\'{\i}}guez-Torres}, {Roe}, {Ross},
  {Ross}, {Rossi}, {Rubi{\~n}o-Mart{\'{\i}}n}, {Saito}, {Salazar-Albornoz},
  {Samushia}, {S{\'a}nchez}, {Satpathy}, {Schlegel}, {Schneider},
  {Sc{\'o}ccola}, {Seo}, {Sheldon}, {Simmons}, {Slosar}, {Strauss}, {Swanson},
  {Thomas}, {Tinker}, {Tojeiro}, {Maga{\~n}a}, {Vazquez}, {Verde}, {Wake},
  {Wang}, {Weinberg}, {White}, {Wood-Vasey}, {Y{\`e}che}, {Zehavi}, {Zhai}, \&
  {Zhao}}]{Alam17}
{Alam}, S., {Ata}, M., {Bailey}, S., {et~al.} 2017, \mnras, 470, 2617

\bibitem[{{Anderson} {et~al.}(2014){Anderson}, {Aubourg}, {Bailey}, {Beutler},
  {Bhardwaj}, {Blanton}, {Bolton}, {Brinkmann}, {Brownstein}, {Burden},
  {Chuang}, {Cuesta}, {Dawson}, {Eisenstein}, {Escoffier}, {Gunn}, {Guo}, {Ho},
  {Honscheid}, {Howlett}, {Kirkby}, {Lupton}, {Manera}, {Maraston}, {McBride},
  {Mena}, {Montesano}, {Nichol}, {Nuza}, {Olmstead}, {Padmanabhan},
  {Palanque-Delabrouille}, {Parejko}, {Percival}, {Petitjean}, {Prada},
  {Price-Whelan}, {Reid}, {Roe}, {Ross}, {Ross}, {Sabiu}, {Saito}, {Samushia},
  {S{\'a}nchez}, {Schlegel}, {Schneider}, {Scoccola}, {Seo}, {Skibba},
  {Strauss}, {Swanson}, {Thomas}, {Tinker}, {Tojeiro}, {Maga{\~n}a}, {Verde},
  {Wake}, {Weaver}, {Weinberg}, {White}, {Xu}, {Y{\`e}che}, {Zehavi}, \&
  {Zhao}}]{Anderson14}
{Anderson}, L., {Aubourg}, {\'E}., {Bailey}, S., {et~al.} 2014, \mnras, 441, 24

\bibitem[{{Argence} \& {Lamareille}(2009)}]{Argence09}
{Argence}, B., \& {Lamareille}, F. 2009, \aap, 495, 759

\bibitem[{{Auger} {et~al.}(2010){Auger}, {Treu}, {Gavazzi}, {Bolton},
  {Koopmans}, \& {Marshall}}]{Auger10}
{Auger}, M.~W., {Treu}, T., {Gavazzi}, R., {et~al.} 2010, \apjl, 721, L163

\bibitem[{Barbary(2014)}]{Barbary14}
Barbary, K. 2014, sncosmo v0.4.2, doi:10.5281/zenodo.11938

\bibitem[{{Barbon} {et~al.}(1979){Barbon}, {Ciatti}, \& {Rosino}}]{Barbon79}
{Barbon}, R., {Ciatti}, F., \& {Rosino}, L. 1979, \aap, 72, 287

\bibitem[{{Bassett} {et~al.}(2006){Bassett}, {Becker}, {Brewington}, {Choi},
  {Cinabro}, {Dejongh}, {Dembicky}, {Depoy}, {Dilday}, {Doi}, {Frieman},
  {Garnavich}, {Harvanek}, {Hogan}, {Holtzman}, {Im}, {Jha}, {Konishi},
  {Krzesinski}, {Lampeitl}, {Kessler}, {Ketzeback}, {Long}, {Malanushenko},
  {Marriner}, {McGinnis}, {McMillan}, {Miknaitis}, {Morokuma}, {Nichol}, {Pan},
  {Prieto}, {Richmond}, {Riess}, {Romani}, {Sako}, {Saurage}, {Schneider},
  {Smith}, {Snedden}, {Takanashi}, {Tokita}, {van der Heyden}, {Watters},
  {Wheeler}, {Yasuda}, {Zheng}, {Hopp}, {Kollatschny}, \&
  {Eastman}}]{Bassett06}
{Bassett}, B., {Becker}, A., {Brewington}, H., {et~al.} 2006, Central Bureau
  Electronic Telegrams, 667

\bibitem[{{Bellm}(2014)}]{ZTF}
{Bellm}, E. 2014, in The Third Hot-wiring the Transient Universe Workshop, ed.
  P.~R. {Wozniak}, M.~J. {Graham}, A.~A. {Mahabal}, \& R.~{Seaman}, 27--33

\bibitem[{{Bennett} {et~al.}(2013){Bennett}, {Larson}, {Weiland}, {Jarosik},
  {Hinshaw}, {Odegard}, {Smith}, {Hill}, {Gold}, {Halpern}, {Komatsu}, {Nolta},
  {Page}, {Spergel}, {Wollack}, {Dunkley}, {Kogut}, {Limon}, {Meyer}, {Tucker},
  \& {Wright}}]{WMAP9}
{Bennett}, C.~L., {Larson}, D., {Weiland}, J.~L., {et~al.} 2013, \apjs, 208, 20

\bibitem[{{Betoule} {et~al.}(2014){Betoule}, {Kessler}, {Guy}, {Mosher},
  {Hardin}, {Biswas}, {Astier}, {El-Hage}, {Konig}, {Kuhlmann}, {Marriner},
  {Pain}, {Regnault}, {Balland}, {Bassett}, {Brown}, {Campbell}, {Carlberg},
  {Cellier-Holzem}, {Cinabro}, {Conley}, {D'Andrea}, {DePoy}, {Doi}, {Ellis},
  {Fabbro}, {Filippenko}, {Foley}, {Frieman}, {Fouchez}, {Galbany}, {Goobar},
  {Gupta}, {Hill}, {Hlozek}, {Hogan}, {Hook}, {Howell}, {Jha}, {Le Guillou},
  {Leloudas}, {Lidman}, {Marshall}, {M{\"o}ller}, {Mour{\~a}o}, {Neveu},
  {Nichol}, {Olmstead}, {Palanque-Delabrouille}, {Perlmutter}, {Prieto},
  {Pritchet}, {Richmond}, {Riess}, {Ruhlmann-Kleider}, {Sako}, {Schahmaneche},
  {Schneider}, {Smith}, {Sollerman}, {Sullivan}, {Walton}, \&
  {Wheeler}}]{Betoule14}
{Betoule}, M., {Kessler}, R., {Guy}, J., {et~al.} 2014, \aap, 568, A22

\bibitem[{{Blandford} \& {Narayan}(1992)}]{Blandford92}
{Blandford}, R.~D., \& {Narayan}, R. 1992, \araa, 30, 311

\bibitem[{{Bolton} \& {Burles}(2003)}]{Bolton03}
{Bolton}, A.~S., \& {Burles}, S. 2003, \apj, 592, 17

\bibitem[{{Bolton} {et~al.}(2008){Bolton}, {Burles}, {Koopmans}, {Treu},
  {Gavazzi}, {Moustakas}, {Wayth}, \& {Schlegel}}]{SLACSV}
{Bolton}, A.~S., {Burles}, S., {Koopmans}, L.~V.~E., {et~al.} 2008, \apj, 682,
  964

\bibitem[{{Bolton} {et~al.}(2006{\natexlab{a}}){Bolton}, {Burles}, {Koopmans},
  {Treu}, \& {Moustakas}}]{SLACSI}
{Bolton}, A.~S., {Burles}, S., {Koopmans}, L.~V.~E., {Treu}, T., \&
  {Moustakas}, L.~A. 2006{\natexlab{a}}, \apj, 638, 703

\bibitem[{{Bolton} {et~al.}(2004){Bolton}, {Burles}, {Schlegel}, {Eisenstein},
  \& {Brinkmann}}]{Bolton04}
{Bolton}, A.~S., {Burles}, S., {Schlegel}, D.~J., {Eisenstein}, D.~J., \&
  {Brinkmann}, J. 2004, \aj, 127, 1860

\bibitem[{{Bolton} {et~al.}(2006{\natexlab{b}}){Bolton}, {Moustakas}, {Stern},
  {Burles}, {Dey}, \& {Spinrad}}]{Bolton06b}
{Bolton}, A.~S., {Moustakas}, L.~A., {Stern}, D., {et~al.} 2006{\natexlab{b}},
  \apjl, 646, L45

\bibitem[{{Bolton} {et~al.}(2012{\natexlab{a}}){Bolton}, {Schlegel}, {Aubourg},
  {Bailey}, {Bhardwaj}, {Brownstein}, {Burles}, {Chen}, {Dawson}, {Eisenstein},
  {Gunn}, {Knapp}, {Loomis}, {Lupton}, {Maraston}, {Muna}, {Myers}, {Olmstead},
  {Padmanabhan}, {P{\^a}ris}, {Percival}, {Petitjean}, {Rockosi}, {Ross},
  {Schneider}, {Shu}, {Strauss}, {Thomas}, {Tremonti}, {Wake}, {Weaver}, \&
  {Wood-Vasey}}]{Bolton12b}
{Bolton}, A.~S., {Schlegel}, D.~J., {Aubourg}, {\'E}., {et~al.}
  2012{\natexlab{a}}, \aj, 144, 144

\bibitem[{{Bolton} {et~al.}(2012{\natexlab{b}}){Bolton}, {Brownstein},
  {Kochanek}, {Shu}, {Schlegel}, {Eisenstein}, {Wake}, {Connolly}, {Maraston},
  {Arneson}, \& {Weaver}}]{Bolton12}
{Bolton}, A.~S., {Brownstein}, J.~R., {Kochanek}, C.~S., {et~al.}
  2012{\natexlab{b}}, \apj, 757, 82

\bibitem[{{Bonvin} {et~al.}(2017){Bonvin}, {Courbin}, {Suyu}, {Marshall},
  {Rusu}, {Sluse}, {Tewes}, {Wong}, {Collett}, {Fassnacht}, {Treu}, {Auger},
  {Hilbert}, {Koopmans}, {Meylan}, {Rumbaugh}, {Sonnenfeld}, \&
  {Spiniello}}]{Bonvin17}
{Bonvin}, V., {Courbin}, F., {Suyu}, S.~H., {et~al.} 2017, \mnras, 465, 4914

\bibitem[{{Brewer} {et~al.}(2014){Brewer}, {Marshall}, {Auger}, {Treu},
  {Dutton}, \& {Barnab{\`e}}}]{Brewer14}
{Brewer}, B.~J., {Marshall}, P.~J., {Auger}, M.~W., {et~al.} 2014, \mnras, 437,
  1950

\bibitem[{{Brinchmann} {et~al.}(2004){Brinchmann}, {Charlot}, {White},
  {Tremonti}, {Kauffmann}, {Heckman}, \& {Brinkmann}}]{Brinchmann04}
{Brinchmann}, J., {Charlot}, S., {White}, S.~D.~M., {et~al.} 2004, \mnras, 351,
  1151

\bibitem[{{Browne} {et~al.}(2003){Browne}, {Wilkinson}, {Jackson}, {Myers},
  {Fassnacht}, {Koopmans}, {Marlow}, {Norbury}, {Rusin}, {Sykes}, {Biggs},
  {Blandford}, {de Bruyn}, {Chae}, {Helbig}, {King}, {McKean}, {Pearson},
  {Phillips}, {Readhead}, {Xanthopoulos}, \& {York}}]{Browne03}
{Browne}, I.~W.~A., {Wilkinson}, P.~N., {Jackson}, N.~J.~F., {et~al.} 2003,
  \mnras, 341, 13

\bibitem[{{Brownstein} {et~al.}(2012){Brownstein}, {Bolton}, {Schlegel},
  {Eisenstein}, {Kochanek}, {Connolly}, {Maraston}, {Pandey}, {Seitz}, {Wake},
  {Wood-Vasey}, {Brinkmann}, {Schneider}, \& {Weaver}}]{Brownstein12}
{Brownstein}, J.~R., {Bolton}, A.~S., {Schlegel}, D.~J., {et~al.} 2012, \apj,
  744, 41

\bibitem[{{Bussmann} {et~al.}(2013){Bussmann}, {P{\'e}rez-Fournon}, {Amber},
  {Calanog}, {Gurwell}, {Dannerbauer}, {De Bernardis}, {Fu}, {Harris}, {Krips},
  {Lapi}, {Maiolino}, {Omont}, {Riechers}, {Wardlow}, {Baker}, {Birkinshaw},
  {Bock}, {Bourne}, {Clements}, {Cooray}, {De Zotti}, {Dunne}, {Dye}, {Eales},
  {Farrah}, {Gavazzi}, {Gonz{\'a}lez Nuevo}, {Hopwood}, {Ibar}, {Ivison},
  {Laporte}, {Maddox}, {Mart{\'{\i}}nez-Navajas}, {Michalowski}, {Negrello},
  {Oliver}, {Roseboom}, {Scott}, {Serjeant}, {Smith}, {Smith}, {Streblyanska},
  {Valiante}, {van der Werf}, {Verma}, {Vieira}, {Wang}, \&
  {Wilner}}]{Bussmann13}
{Bussmann}, R.~S., {P{\'e}rez-Fournon}, I., {Amber}, S., {et~al.} 2013, \apj,
  779, 25

\bibitem[{{Calzetti} {et~al.}(2000){Calzetti}, {Armus}, {Bohlin}, {Kinney},
  {Koornneef}, \& {Storchi-Bergmann}}]{Calzetti2000}
{Calzetti}, D., {Armus}, L., {Bohlin}, R.~C., {et~al.} 2000, \apj, 533, 682

\bibitem[{{Cao} {et~al.}(2013){Cao}, {Kasliwal}, {Arcavi}, {Horesh}, {Hancock},
  {Valenti}, {Cenko}, {Kulkarni}, {Gal-Yam}, {Gorbikov}, {Ofek}, {Sand},
  {Yaron}, {Graham}, {Silverman}, {Wheeler}, {Marion}, {Walker}, {Mazzali},
  {Howell}, {Li}, {Kong}, {Bloom}, {Nugent}, {Surace}, {Masci}, {Carpenter},
  {Degenaar}, \& {Gelino}}]{Cao13}
{Cao}, Y., {Kasliwal}, M.~M., {Arcavi}, I., {et~al.} 2013, \apjl, 775, L7

\bibitem[{{Cappellari} {et~al.}(2012){Cappellari}, {McDermid}, {Alatalo},
  {Blitz}, {Bois}, {Bournaud}, {Bureau}, {Crocker}, {Davies}, {Davis}, {de
  Zeeuw}, {Duc}, {Emsellem}, {Khochfar}, {Krajnovi{\'c}}, {Kuntschner},
  {Lablanche}, {Morganti}, {Naab}, {Oosterloo}, {Sarzi}, {Scott}, {Serra},
  {Weijmans}, \& {Young}}]{Cappellari12}
{Cappellari}, M., {McDermid}, R.~M., {Alatalo}, K., {et~al.} 2012, \nat, 484,
  485

\bibitem[{{Cardiel} {et~al.}(2003){Cardiel}, {Elbaz}, {Schiavon}, {Willmer},
  {Koo}, {Phillips}, \& {Gallego}}]{Cardiel03}
{Cardiel}, N., {Elbaz}, D., {Schiavon}, R.~P., {et~al.} 2003, \apj, 584, 76

\bibitem[{{Christensen} {et~al.}(2012){Christensen}, {Richard}, {Hjorth},
  {Milvang-Jensen}, {Laursen}, {Limousin}, {Dessauges-Zavadsky}, {Grillo}, \&
  {Ebeling}}]{Christensen12}
{Christensen}, L., {Richard}, J., {Hjorth}, J., {et~al.} 2012, \mnras, 427,
  1953

\bibitem[{{Collett}(2015)}]{Collett15}
{Collett}, T.~E. 2015, \apj, 811, 20

\bibitem[{{Conroy} {et~al.}(2013){Conroy}, {Dutton}, {Graves}, {Mendel}, \&
  {van Dokkum}}]{Conroy13}
{Conroy}, C., {Dutton}, A.~A., {Graves}, G.~J., {Mendel}, J.~T., \& {van
  Dokkum}, P.~G. 2013, \apjl, 776, L26

\bibitem[{{Dahl{\'e}n} \& {Fransson}(1999)}]{Dahlen99}
{Dahl{\'e}n}, T., \& {Fransson}, C. 1999, \aap, 350, 349

\bibitem[{{Dahlen} {et~al.}(2012){Dahlen}, {Strolger}, {Riess}, {Mattila},
  {Kankare}, \& {Mobasher}}]{Dahlen12}
{Dahlen}, T., {Strolger}, L.-G., {Riess}, A.~G., {et~al.} 2012, \apj, 757, 70

\bibitem[{{Dahlen} {et~al.}(2004){Dahlen}, {Strolger}, {Riess}, {Mobasher},
  {Chary}, {Conselice}, {Ferguson}, {Fruchter}, {Giavalisco}, {Livio}, {Madau},
  {Panagia}, \& {Tonry}}]{Dahlen04}
---. 2004, \apj, 613, 189

\bibitem[{{de Jaeger} {et~al.}(2017){de Jaeger}, {Gonz{\'a}lez-Gait{\'a}n},
  {Hamuy}, {Galbany}, {Anderson}, {Phillips}, {Stritzinger}, {Carlberg},
  {Sullivan}, {Guti{\'e}rrez}, {Hook}, {Howell}, {Hsiao}, {Kuncarayakti},
  {Ruhlmann-Kleider}, {Folatelli}, {Pritchet}, \& {Basa}}]{deJaeger17}
{de Jaeger}, T., {Gonz{\'a}lez-Gait{\'a}n}, S., {Hamuy}, M., {et~al.} 2017,
  \apj, 835, 166

\bibitem[{{Diehl} {et~al.}(2017){Diehl}, {Buckley-Geer}, {Lindgren}, {Nord},
  {Gaitsch}, {Gaitsch}, {Lin}, {Allam}, {Collett}, {Furlanetto}, {Gill},
  {More}, {Nightingale}, {Odden}, {Pellico}, {Tucker}, {da Costa}, {Fausti
  Neto}, {Kuropatkin}, {Soares-Santos}, {Welch}, {Zhang}, {Frieman}, {Abdalla},
  {Annis}, {Benoit-L{\'e}vy}, {Bertin}, {Brooks}, {Burke}, {Carnero Rosell},
  {Carrasco Kind}, {Carretero}, {Cunha}, {D'Andrea}, {Desai}, {Dietrich},
  {Drlica-Wagner}, {Evrard}, {Finley}, {Flaugher}, {Garc{\'{\i}}a-Bellido},
  {Gerdes}, {Goldstein}, {Gruen}, {Gruendl}, {Gschwend}, {Gutierrez}, {James},
  {Kuehn}, {Kuhlmann}, {Lahav}, {Li}, {Lima}, {Maia}, {Marshall}, {Menanteau},
  {Miquel}, {Nichol}, {Nugent}, {Ogando}, {Plazas}, {Reil}, {Romer}, {Sako},
  {Sanchez}, {Santiago}, {Scarpine}, {Schindler}, {Schubnell},
  {Sevilla-Noarbe}, {Sheldon}, {Smith}, {Sobreira}, {Suchyta}, {Swanson},
  {Tarle}, {Thomas}, {Walker}, \& {DES Collaboration}}]{Diehl17}
{Diehl}, H.~T., {Buckley-Geer}, E.~J., {Lindgren}, K.~A., {et~al.} 2017, \apjs,
  232, 15

\bibitem[{{Diehl} {et~al.}(2006){Diehl}, {Halloin}, {Kretschmer}, {Lichti},
  {Sch{\"o}nfelder}, {Strong}, {von Kienlin}, {Wang}, {Jean}, {Kn{\"o}dlseder},
  {Roques}, {Weidenspointner}, {Schanne}, {Hartmann}, {Winkler}, \&
  {Wunderer}}]{Diehl06}
{Diehl}, R., {Halloin}, H., {Kretschmer}, K., {et~al.} 2006, \nat, 439, 45

\bibitem[{{Dilday} {et~al.}(2012){Dilday}, {Howell}, {Cenko}, {Silverman},
  {Nugent}, {Sullivan}, {Ben-Ami}, {Bildsten}, {Bolte}, {Endl}, {Filippenko},
  {Gnat}, {Horesh}, {Hsiao}, {Kasliwal}, {Kirkman}, {Maguire}, {Marcy},
  {Moore}, {Pan}, {Parrent}, {Podsiadlowski}, {Quimby}, {Sternberg}, {Suzuki},
  {Tytler}, {Xu}, {Bloom}, {Gal-Yam}, {Hook}, {Kulkarni}, {Law}, {Ofek},
  {Polishook}, \& {Poznanski}}]{Dilday12}
{Dilday}, B., {Howell}, D.~A., {Cenko}, S.~B., {et~al.} 2012, Science, 337, 942

\bibitem[{{Dobler} \& {Keeton}(2006)}]{Dobler06}
{Dobler}, G., \& {Keeton}, C.~R. 2006, \apj, 653, 1391

\bibitem[{{Doggett} \& {Branch}(1985)}]{Doggett85}
{Doggett}, J.~B., \& {Branch}, D. 1985, \aj, 90, 2303

\bibitem[{{Doi} {et~al.}(2010){Doi}, {Tanaka}, {Fukugita}, {Gunn}, {Yasuda},
  {Ivezi{\'c}}, {Brinkmann}, {de Haars}, {Kleinman}, {Krzesinski}, \& {French
  Leger}}]{Doi10}
{Doi}, M., {Tanaka}, M., {Fukugita}, M., {et~al.} 2010, \aj, 139, 1628

\bibitem[{{Dubois} {et~al.}(2013){Dubois}, {Gavazzi}, {Peirani}, \&
  {Silk}}]{Dubois13}
{Dubois}, Y., {Gavazzi}, R., {Peirani}, S., \& {Silk}, J. 2013, \mnras, 433,
  3297

\bibitem[{{Duffy} {et~al.}(2010){Duffy}, {Schaye}, {Kay}, {Dalla Vecchia},
  {Battye}, \& {Booth}}]{Duffy10}
{Duffy}, A.~R., {Schaye}, J., {Kay}, S.~T., {et~al.} 2010, \mnras, 405, 2161

\bibitem[{{Eisenstein} {et~al.}(2005){Eisenstein}, {Zehavi}, {Hogg},
  {Scoccimarro}, {Blanton}, {Nichol}, {Scranton}, {Seo}, {Tegmark}, {Zheng},
  {Anderson}, {Annis}, {Bahcall}, {Brinkmann}, {Burles}, {Castander},
  {Connolly}, {Csabai}, {Doi}, {Fukugita}, {Frieman}, {Glazebrook}, {Gunn},
  {Hendry}, {Hennessy}, {Ivezi{\'c}}, {Kent}, {Knapp}, {Lin}, {Loh}, {Lupton},
  {Margon}, {McKay}, {Meiksin}, {Munn}, {Pope}, {Richmond}, {Schlegel},
  {Schneider}, {Shimasaku}, {Stoughton}, {Strauss}, {SubbaRao}, {Szalay},
  {Szapudi}, {Tucker}, {Yanny}, \& {York}}]{Eisenstein05}
{Eisenstein}, D.~J., {Zehavi}, I., {Hogg}, D.~W., {et~al.} 2005, \apj, 633, 560

\bibitem[{{Falco} {et~al.}(1985){Falco}, {Gorenstein}, \& {Shapiro}}]{Falco85}
{Falco}, E.~E., {Gorenstein}, M.~V., \& {Shapiro}, I.~I. 1985, \apjl, 289, L1

\bibitem[{{Faure} {et~al.}(2008){Faure}, {Kneib}, {Covone}, {Tasca},
  {Leauthaud}, {Capak}, {Jahnke}, {Smolcic}, {de la Torre}, {Ellis},
  {Finoguenov}, {Koekemoer}, {Le Fevre}, {Massey}, {Mellier}, {Refregier},
  {Rhodes}, {Scoville}, {Schinnerer}, {Taylor}, {Van Waerbeke}, \&
  {Walcher}}]{Faure08}
{Faure}, C., {Kneib}, J.-P., {Covone}, G., {et~al.} 2008, \apjs, 176, 19

\bibitem[{{Ferreras} {et~al.}(2013){Ferreras}, {La Barbera}, {de la Rosa},
  {Vazdekis}, {de Carvalho}, {Falc{\'o}n-Barroso}, \&
  {Ricciardelli}}]{Ferreras13}
{Ferreras}, I., {La Barbera}, F., {de la Rosa}, I.~G., {et~al.} 2013, \mnras,
  429, L15

\bibitem[{{Firth} {et~al.}(2015){Firth}, {Sullivan}, {Gal-Yam}, {Howell},
  {Maguire}, {Nugent}, {Piro}, {Baltay}, {Feindt}, {Hadjiyksta}, {McKinnon},
  {Ofek}, {Rabinowitz}, \& {Walker}}]{Firth15}
{Firth}, R.~E., {Sullivan}, M., {Gal-Yam}, A., {et~al.} 2015, \mnras, 446, 3895

\bibitem[{{Foley} {et~al.}(2015){Foley}, {Van Dyk}, {Jha}, {Clubb},
  {Filippenko}, {Mauerhan}, {Miller}, \& {Smith}}]{Foley15}
{Foley}, R.~J., {Van Dyk}, S.~D., {Jha}, S.~W., {et~al.} 2015, \apjl, 798, L37

\bibitem[{{Foxley-Marrable} {et~al.}(2018){Foxley-Marrable}, {Collett},
  {Vernardos}, {Goldstein}, \& {Bacon}}]{Foxley-Marrable18}
{Foxley-Marrable}, M., {Collett}, T.~E., {Vernardos}, G., {Goldstein}, D.~A.,
  \& {Bacon}, D. 2018, ArXiv e-prints, arXiv:1802.07738

\bibitem[{{Gal-Yam} {et~al.}(2011){Gal-Yam}, {Kasliwal}, {Arcavi}, {Green},
  {Yaron}, {Ben-Ami}, {Xu}, {Sternberg}, {Quimby}, {Kulkarni}, {Ofek},
  {Walters}, {Nugent}, {Poznanski}, {Bloom}, {Cenko}, {Filippenko}, {Li},
  {Silverman}, {Walker}, {Sullivan}, {Maguire}, {Howell}, {Mazzali}, {Frail},
  {Bersier}, {James}, {Akerlof}, {Yuan}, {Law}, {Fox}, \&
  {Gehrels}}]{Gal-Yam11}
{Gal-Yam}, A., {Kasliwal}, M.~M., {Arcavi}, I., {et~al.} 2011, \apj, 736, 159

\bibitem[{{Gal-Yam} {et~al.}(2014){Gal-Yam}, {Arcavi}, {Ofek}, {Ben-Ami},
  {Cenko}, {Kasliwal}, {Cao}, {Yaron}, {Tal}, {Silverman}, {Horesh}, {De Cia},
  {Taddia}, {Sollerman}, {Perley}, {Vreeswijk}, {Kulkarni}, {Nugent},
  {Filippenko}, \& {Wheeler}}]{Gal-Yam14}
{Gal-Yam}, A., {Arcavi}, I., {Ofek}, E.~O., {et~al.} 2014, \nat, 509, 471

\bibitem[{{Gallagher} {et~al.}(1989){Gallagher}, {Hunter}, \&
  {Bushouse}}]{Gallagher89}
{Gallagher}, J.~S., {Hunter}, D.~A., \& {Bushouse}, H. 1989, \aj, 97, 700

\bibitem[{{Ganot} {et~al.}(2016){Ganot}, {Gal-Yam}, {Ofek}, {Sagiv}, {Waxman},
  {Lapid}, {Kulkarni}, {Ben-Ami}, {Kasliwal}, {The ULTRASAT Science Team},
  {Chelouche}, {Rafter}, {Behar}, {Laor}, {Poznanski}, {Nakar}, {Maoz},
  {Trakhtenbrot}, {WTTH Consortium}, {Neill}, {Barlow}, {Martin}, {Gezari},
  {the GALEX Science Team}, {Arcavi}, {Bloom}, {Nugent}, {Sullivan}, \&
  {Palomar Transient Factory}}]{Ganot16}
{Ganot}, N., {Gal-Yam}, A., {Ofek}, E.~O., {et~al.} 2016, \apj, 820, 57

\bibitem[{{Gilbank} {et~al.}(2010){Gilbank}, {Baldry}, {Balogh}, {Glazebrook},
  \& {Bower}}]{Gilbank10}
{Gilbank}, D.~G., {Baldry}, I.~K., {Balogh}, M.~L., {Glazebrook}, K., \&
  {Bower}, R.~G. 2010, \mnras, 405, 2594

\bibitem[{{Gnedin} {et~al.}(2004){Gnedin}, {Kravtsov}, {Klypin}, \&
  {Nagai}}]{Gnedin04}
{Gnedin}, O.~Y., {Kravtsov}, A.~V., {Klypin}, A.~A., \& {Nagai}, D. 2004, \apj,
  616, 16

\bibitem[{{Goldstein} \& {Nugent}(2017)}]{Goldstein17}
{Goldstein}, D.~A., \& {Nugent}, P.~E. 2017, \apjl, 834, L5

\bibitem[{{Goldstein} {et~al.}(2018){Goldstein}, {Nugent}, {Kasen}, \&
  {Collett}}]{Goldstein17b}
{Goldstein}, D.~A., {Nugent}, P.~E., {Kasen}, D.~N., \& {Collett}, T.~E. 2018,
  \apj, 855, 22

\bibitem[{{Goobar} {et~al.}(2009){Goobar}, {Paech}, {Stanishev}, {Amanullah},
  {Dahl{\'e}n}, {J{\"o}nsson}, {Kneib}, {Lidman}, {Limousin}, {M{\"o}rtsell},
  {Nobili}, {Richard}, {Riehm}, \& {von Strauss}}]{Goobar09}
{Goobar}, A., {Paech}, K., {Stanishev}, V., {et~al.} 2009, \aap, 507, 71

\bibitem[{{Goobar} {et~al.}(2017){Goobar}, {Amanullah}, {Kulkarni}, {Nugent},
  {Johansson}, {Steidel}, {Law}, {M{\"o}rtsell}, {Quimby}, {Blagorodnova},
  {Brandeker}, {Cao}, {Cooray}, {Ferretti}, {Fremling}, {Hangard}, {Kasliwal},
  {Kupfer}, {Lunnan}, {Masci}, {Miller}, {Nayyeri}, {Neill}, {Ofek},
  {Papadogiannakis}, {Petrushevska}, {Ravi}, {Sollerman}, {Sullivan}, {Taddia},
  {Walters}, {Wilson}, {Yan}, \& {Yaron}}]{Goobar17}
{Goobar}, A., {Amanullah}, R., {Kulkarni}, S.~R., {et~al.} 2017, Science, 356,
  291

\bibitem[{{Gorenstein} {et~al.}(1988){Gorenstein}, {Shapiro}, \&
  {Falco}}]{Gorenstein88}
{Gorenstein}, M.~V., {Shapiro}, I.~I., \& {Falco}, E.~E. 1988, \apj, 327, 693

\bibitem[{{Governato} {et~al.}(2010){Governato}, {Brook}, {Mayer}, {Brooks},
  {Rhee}, {Wadsley}, {Jonsson}, {Willman}, {Stinson}, {Quinn}, \&
  {Madau}}]{Governato10}
{Governato}, F., {Brook}, C., {Mayer}, L., {et~al.} 2010, \nat, 463, 203

\bibitem[{{Graur} {et~al.}(2015){Graur}, {Bianco}, \& {Modjaz}}]{Graur15}
{Graur}, O., {Bianco}, F.~B., \& {Modjaz}, M. 2015, \mnras, 450, 905

\bibitem[{{Greggio} {et~al.}(2008){Greggio}, {Renzini}, \& {Daddi}}]{Greggio08}
{Greggio}, L., {Renzini}, A., \& {Daddi}, E. 2008, \mnras, 388, 829

\bibitem[{{Gustafsson} {et~al.}(2006){Gustafsson}, {Fairbairn}, \&
  {Sommer-Larsen}}]{Gustafsson06}
{Gustafsson}, M., {Fairbairn}, M., \& {Sommer-Larsen}, J. 2006, \prd, 74,
  123522

\bibitem[{{Hayashi} {et~al.}(2013){Hayashi}, {Sobral}, {Best}, {Smail}, \&
  {Kodama}}]{Hayashi13}
{Hayashi}, M., {Sobral}, D., {Best}, P.~N., {Smail}, I., \& {Kodama}, T. 2013,
  \mnras, 430, 1042

\bibitem[{{He} {et~al.}(2018){He}, {Wang}, \& {Huang}}]{He18}
{He}, S., {Wang}, L., \& {Huang}, J.~Z. 2018, ArXiv e-prints, arXiv:1802.06125

\bibitem[{{Hicks} {et~al.}(2002){Hicks}, {Malkan}, {Teplitz}, {McCarthy}, \&
  {Yan}}]{Hicks02}
{Hicks}, E.~K.~S., {Malkan}, M.~A., {Teplitz}, H.~I., {McCarthy}, P.~J., \&
  {Yan}, L. 2002, \apj, 581, 205

\bibitem[{{Hoeflich} {et~al.}(2017){Hoeflich}, {Hsiao}, {Ashall}, {Burns},
  {Diamond}, {Phillips}, {Sand}, {Stritzinger}, {Suntzeff}, {Contreras},
  {Krisciunas}, {Morrell}, \& {Wang}}]{Hoeflich17}
{Hoeflich}, P., {Hsiao}, E.~Y., {Ashall}, C., {et~al.} 2017, \apj, 846, 58

\bibitem[{{Hogg} {et~al.}(1998){Hogg}, {Cohen}, {Blandford}, \&
  {Pahre}}]{Hogg98}
{Hogg}, D.~W., {Cohen}, J.~G., {Blandford}, R., \& {Pahre}, M.~A. 1998, \apj,
  504, 622

\bibitem[{{Hopkins} \& {Beacom}(2006)}]{Hopkins06}
{Hopkins}, A.~M., \& {Beacom}, J.~F. 2006, \apj, 651, 142

\bibitem[{{Hopkins} {et~al.}(2003){Hopkins}, {Miller}, {Nichol}, {Connolly},
  {Bernardi}, {G{\'o}mez}, {Goto}, {Tremonti}, {Brinkmann}, {Ivezi{\'c}}, \&
  {Lamb}}]{Hopkins03}
{Hopkins}, A.~M., {Miller}, C.~J., {Nichol}, R.~C., {et~al.} 2003, \apj, 599,
  971

\bibitem[{{Hsiao} {et~al.}(2007){Hsiao}, {Conley}, {Howell}, {Sullivan},
  {Pritchet}, {Carlberg}, {Nugent}, \& {Phillips}}]{Hsiao07}
{Hsiao}, E.~Y., {Conley}, A., {Howell}, D.~A., {et~al.} 2007, \apj, 663, 1187

\bibitem[{{Impey} {et~al.}(1998){Impey}, {Falco}, {Kochanek}, {Leh{\'a}r},
  {McLeod}, {Rix}, {Peng}, \& {Keeton}}]{Impey98}
{Impey}, C.~D., {Falco}, E.~E., {Kochanek}, C.~S., {et~al.} 1998, \apj, 509,
  551

\bibitem[{{Inada} {et~al.}(2012){Inada}, {Oguri}, {Shin}, {Kayo}, {Strauss},
  {Morokuma}, {Rusu}, {Fukugita}, {Kochanek}, {Richards}, {Schneider}, {York},
  {Bahcall}, {Frieman}, {Hall}, \& {White}}]{Inada12}
{Inada}, N., {Oguri}, M., {Shin}, M.-S., {et~al.} 2012, \aj, 143, 119

\bibitem[{{Ivezic} {et~al.}(2008){Ivezic}, {Tyson}, {Abel}, {Acosta},
  {Allsman}, {AlSayyad}, {Anderson}, {Andrew}, {Angel}, {Angeli}, {Ansari},
  {Antilogus}, {Arndt}, {Astier}, {Aubourg}, {Axelrod}, {Bard}, {Barr},
  {Barrau}, {Bartlett}, {Bauman}, {Beaumont}, {Becker}, {Becla}, {Beldica},
  {Bellavia}, {Blanc}, {Blandford}, {Bloom}, {Bogart}, {Borne}, {Bosch},
  {Boutigny}, {Brandt}, {Brown}, {Bullock}, {Burchat}, {Burke}, {Cagnoli},
  {Calabrese}, {Chandrasekharan}, {Chesley}, {Cheu}, {Chiang}, {Claver},
  {Connolly}, {Cook}, {Cooray}, {Covey}, {Cribbs}, {Cui}, {Cutri}, {Daubard},
  {Daues}, {Delgado}, {Digel}, {Doherty}, {Dubois}, {Dubois-Felsmann},
  {Durech}, {Eracleous}, {Ferguson}, {Frank}, {Freemon}, {Gangler}, {Gawiser},
  {Geary}, {Gee}, {Geha}, {Gibson}, {Gilmore}, {Glanzman}, {Goodenow},
  {Gressler}, {Gris}, {Guyonnet}, {Hascall}, {Haupt}, {Hernandez}, {Hogan},
  {Huang}, {Huffer}, {Innes}, {Jacoby}, {Jain}, {Jee}, {Jernigan},
  {Jevremovic}, {Johns}, {Jones}, {Juramy-Gilles}, {Juric}, {Kahn}, {Kalirai},
  {Kallivayalil}, {Kalmbach}, {Kantor}, {Kasliwal}, {Kessler}, {Kirkby},
  {Knox}, {Kotov}, {Krabbendam}, {Krughoff}, {Kubanek}, {Kuczewski},
  {Kulkarni}, {Lambert}, {Le Guillou}, {Levine}, {Liang}, {Lim}, {Lintott},
  {Lupton}, {Mahabal}, {Marshall}, {Marshall}, {May}, {McKercher}, {Migliore},
  {Miller}, {Mills}, {Monet}, {Moniez}, {Neill}, {Nief}, {Nomerotski},
  {Nordby}, {O'Connor}, {Oliver}, {Olivier}, {Olsen}, {Ortiz}, {Owen}, {Pain},
  {Peterson}, {Petry}, {Pierfederici}, {Pietrowicz}, {Pike}, {Pinto}, {Plante},
  {Plate}, {Price}, {Prouza}, {Radeka}, {Rajagopal}, {Rasmussen}, {Regnault},
  {Ridgway}, {Ritz}, {Rosing}, {Roucelle}, {Rumore}, {Russo}, {Saha},
  {Sassolas}, {Schalk}, {Schindler}, {Schneider}, {Schumacher}, {Sebag},
  {Sembroski}, {Seppala}, {Shipsey}, {Silvestri}, {Smith}, {Smith}, {Strauss},
  {Stubbs}, {Sweeney}, {Szalay}, {Takacs}, {Thaler}, {Van Berg}, {Vanden Berk},
  {Vetter}, {Virieux}, {Xin}, {Walkowicz}, {Walter}, {Wang}, {Warner},
  {Willman}, {Wittman}, {Wolff}, {Wood-Vasey}, {Yoachim}, {Zhan}, \& {for the
  LSST Collaboration}}]{LSST}
{Ivezic}, Z., {Tyson}, J.~A., {Abel}, B., {et~al.} 2008, ArXiv e-prints,
  arXiv:0805.2366

\bibitem[{{Jansen} {et~al.}(2001){Jansen}, {Franx}, \& {Fabricant}}]{Jansen01}
{Jansen}, R.~A., {Franx}, M., \& {Fabricant}, D. 2001, \apj, 551, 825

\bibitem[{{Jones} {et~al.}(2009){Jones}, {Hamuy}, {Lira}, {Maza},
  {Clocchiatti}, {Phillips}, {Morrell}, {Roth}, {Suntzeff}, {Matheson},
  {Filippenko}, {Foley}, \& {Leonard}}]{Jones09}
{Jones}, M.~I., {Hamuy}, M., {Lira}, P., {et~al.} 2009, \apj, 696, 1176

\bibitem[{{Joubert} {et~al.}(2006){Joubert}, {Li}, {Itagaki}, \&
  {Nakano}}]{Joubert06}
{Joubert}, N., {Li}, W., {Itagaki}, K., \& {Nakano}, S. 2006, Central Bureau
  Electronic Telegrams, 609

\bibitem[{{Karman} {et~al.}(2016){Karman}, {Grillo}, {Balestra}, {Rosati},
  {Caputi}, {Di Teodoro}, {Fraternali}, {Gavazzi}, {Mercurio}, {Prochaska},
  {Rodney}, \& {Treu}}]{Karman16}
{Karman}, W., {Grillo}, C., {Balestra}, I., {et~al.} 2016, \aap, 585, A27

\bibitem[{{Kasen} \& {Woosley}(2009)}]{Kasen09}
{Kasen}, D., \& {Woosley}, S.~E. 2009, \apj, 703, 2205

\bibitem[{{Kelly} {et~al.}(2014){Kelly}, {Fox}, {Filippenko}, {Cenko}, {Prato},
  {Schaefer}, {Shen}, {Zheng}, {Graham}, \& {Tucker}}]{Kelly14}
{Kelly}, P.~L., {Fox}, O.~D., {Filippenko}, A.~V., {et~al.} 2014, \apj, 790, 3

\bibitem[{{Kelly} {et~al.}(2015){Kelly}, {Rodney}, {Treu}, {Foley}, {Brammer},
  {Schmidt}, {Zitrin}, {Sonnenfeld}, {Strolger}, {Graur}, {Filippenko}, {Jha},
  {Riess}, {Bradac}, {Weiner}, {Scolnic}, {Malkan}, {von der Linden}, {Trenti},
  {Hjorth}, {Gavazzi}, {Fontana}, {Merten}, {McCully}, {Jones}, {Postman},
  {Dressler}, {Patel}, {Cenko}, {Graham}, \& {Tucker}}]{Kelly15}
{Kelly}, P.~L., {Rodney}, S.~A., {Treu}, T., {et~al.} 2015, Science, 347, 1123

\bibitem[{{Kelly} {et~al.}(2016){Kelly}, {Rodney}, {Treu}, {Strolger}, {Foley},
  {Jha}, {Selsing}, {Brammer}, {Brada{\v c}}, {Cenko}, {Graur}, {Filippenko},
  {Hjorth}, {McCully}, {Molino}, {Nonino}, {Riess}, {Schmidt}, {Tucker}, {von
  der Linden}, {Weiner}, \& {Zitrin}}]{Kelly16}
---. 2016, \apjl, 819, L8

\bibitem[{{Kennicutt}(1983)}]{Kennicutt93}
{Kennicutt}, Jr., R.~C. 1983, \apj, 272, 54

\bibitem[{{Kennicutt}(1992)}]{Kennicutt92}
---. 1992, \apj, 388, 310

\bibitem[{{Kennicutt}(1998)}]{Kennicutt98}
---. 1998, \araa, 36, 189

\bibitem[{{Kewley} {et~al.}(2004){Kewley}, {Geller}, \& {Jansen}}]{Kewley04}
{Kewley}, L.~J., {Geller}, M.~J., \& {Jansen}, R.~A. 2004, \aj, 127, 2002

\bibitem[{{Koopmans} {et~al.}(2003){Koopmans}, {Treu}, {Fassnacht},
  {Blandford}, \& {Surpi}}]{Koopmans03}
{Koopmans}, L.~V.~E., {Treu}, T., {Fassnacht}, C.~D., {Blandford}, R.~D., \&
  {Surpi}, G. 2003, \apj, 599, 70

\bibitem[{{Kostrzewa-Rutkowska} {et~al.}(2013){Kostrzewa-Rutkowska},
  {Wyrzykowski}, \& {Jaroszy{\'n}ski}}]{Kostrzewa-Rutkowska13}
{Kostrzewa-Rutkowska}, Z., {Wyrzykowski}, {\L}., \& {Jaroszy{\'n}ski}, M. 2013,
  \mnras, 429, 2392

\bibitem[{{Kroupa}(2001)}]{Kroupa01}
{Kroupa}, P. 2001, \mnras, 322, 231

\bibitem[{{La Barbera} {et~al.}(2013){La Barbera}, {Ferreras}, {Vazdekis}, {de
  la Rosa}, {de Carvalho}, {Trevisan}, {Falc{\'o}n-Barroso}, \&
  {Ricciardelli}}]{LaBarbera13}
{La Barbera}, F., {Ferreras}, I., {Vazdekis}, A., {et~al.} 2013, \mnras, 433,
  3017

\bibitem[{{Li} {et~al.}(2017){Li}, {Ge}, {Mao}, {Cappellari}, {Long}, {Li},
  {Emsellem}, {Dutton}, {Li}, {Bundy}, {Thomas}, {Drory}, \& {Lopes}}]{Li17}
{Li}, H., {Ge}, J., {Mao}, S., {et~al.} 2017, \apj, 838, 77

\bibitem[{{Li} {et~al.}(2011{\natexlab{a}}){Li}, {Chornock}, {Leaman},
  {Filippenko}, {Poznanski}, {Wang}, {Ganeshalingam}, \& {Mannucci}}]{Li11c}
{Li}, W., {Chornock}, R., {Leaman}, J., {et~al.} 2011{\natexlab{a}}, \mnras,
  412, 1473

\bibitem[{{Li} {et~al.}(2011{\natexlab{b}}){Li}, {Bloom}, {Podsiadlowski},
  {Miller}, {Cenko}, {Jha}, {Sullivan}, {Howell}, {Nugent}, {Butler}, {Ofek},
  {Kasliwal}, {Richards}, {Stockton}, {Shih}, {Bildsten}, {Shara}, {Bibby},
  {Filippenko}, {Ganeshalingam}, {Silverman}, {Kulkarni}, {Law}, {Poznanski},
  {Quimby}, {McCully}, {Patel}, {Maguire}, \& {Shen}}]{Li11}
{Li}, W., {Bloom}, J.~S., {Podsiadlowski}, P., {et~al.} 2011{\natexlab{b}},
  \nat, 480, 348

\bibitem[{{Li} {et~al.}(2011{\natexlab{c}}){Li}, {Leaman}, {Chornock},
  {Filippenko}, {Poznanski}, {Ganeshalingam}, {Wang}, {Modjaz}, {Jha}, {Foley},
  \& {Smith}}]{Li11b}
{Li}, W., {Leaman}, J., {Chornock}, R., {et~al.} 2011{\natexlab{c}}, \mnras,
  412, 1441

\bibitem[{{Liesenborgs} \& {De Rijcke}(2012)}]{Liesenborgs12}
{Liesenborgs}, J., \& {De Rijcke}, S. 2012, \mnras, 425, 1772

\bibitem[{{LSST Science Collaboration} {et~al.}(2017){LSST Science
  Collaboration}, {Marshall}, {Anguita}, {Bianco}, {Bellm}, {Brandt},
  {Clarkson}, {Connolly}, {Gawiser}, {Ivezic}, {Jones}, {Lochner}, {Lund},
  {Mahabal}, {Nidever}, {Olsen}, {Ridgway}, {Rhodes}, {Shemmer}, {Trilling},
  {Vivas}, {Walkowicz}, {Willman}, {Yoachim}, {Anderson}, {Antilogus}, {Angus},
  {Arcavi}, {Awan}, {Biswas}, {Bell}, {Bennett}, {Britt}, {Buzasi},
  {Casetti-Dinescu}, {Chomiuk}, {Claver}, {Cook}, {Davenport}, {Debattista},
  {Digel}, {Doctor}, {Firth}, {Foley}, {Fong}, {Galbany}, {Giampapa}, {Gizis},
  {Graham}, {Grillmair}, {Gris}, {Haiman}, {Hartigan}, {Hawley}, {Hlozek},
  {Jha}, {Johns-Krull}, {Kanbur}, {Kalogera}, {Kashyap}, {Kasliwal}, {Kessler},
  {Kim}, {Kurczynski}, {Lahav}, {Liu}, {Malz}, {Margutti}, {Matheson},
  {McEwen}, {McGehee}, {Meibom}, {Meyers}, {Monet}, {Neilsen}, {Newman},
  {O'Dowd}, {Peiris}, {Penny}, {Peters}, {Poleski}, {Ponder}, {Richards},
  {Rho}, {Rubin}, {Schmidt}, {Schuhmann}, {Shporer}, {Slater}, {Smith},
  {Soares-Santos}, {Stassun}, {Strader}, {Strauss}, {Street}, {Stubbs},
  {Sullivan}, {Szkody}, {Trimble}, {Tyson}, {de Val-Borro}, {Valenti},
  {Wagoner}, {Wood-Vasey}, \& {Zauderer}}]{LSST17}
{LSST Science Collaboration}, {Marshall}, P., {Anguita}, T., {et~al.} 2017,
  ArXiv e-prints, arXiv:1708.04058

\bibitem[{{Madau} {et~al.}(1998){Madau}, {della Valle}, \& {Panagia}}]{Madau98}
{Madau}, P., {della Valle}, M., \& {Panagia}, N. 1998, \mnras, 297, L17

\bibitem[{{Madau} \& {Dickinson}(2014)}]{Madau14}
{Madau}, P., \& {Dickinson}, M. 2014, \araa, 52, 415

\bibitem[{{Mannucci} {et~al.}(2003){Mannucci}, {Maiolino}, {Cresci}, {Della
  Valle}, {Vanzi}, {Ghinassi}, {Ivanov}, {Nagar}, \&
  {Alonso-Herrero}}]{Mannucci03}
{Mannucci}, F., {Maiolino}, R., {Cresci}, G., {et~al.} 2003, \aap, 401, 519

\bibitem[{{Maoz} {et~al.}(2012){Maoz}, {Mannucci}, \& {Brandt}}]{Maoz12}
{Maoz}, D., {Mannucci}, F., \& {Brandt}, T.~D. 2012, \mnras, 426, 3282

\bibitem[{{Maoz} {et~al.}(2011){Maoz}, {Mannucci}, {Li}, {Filippenko}, {Della
  Valle}, \& {Panagia}}]{Maoz11}
{Maoz}, D., {Mannucci}, F., {Li}, W., {et~al.} 2011, \mnras, 412, 1508

\bibitem[{{Markwardt}(2009)}]{MPFIT}
{Markwardt}, C.~B. 2009, in Astronomical Society of the Pacific Conference
  Series, Vol. 411, Astronomical Data Analysis Software and Systems XVIII, ed.
  D.~A. {Bohlender}, D.~{Durand}, \& P.~{Dowler}, 251

\bibitem[{{Marshall} {et~al.}(2009){Marshall}, {Hogg}, {Moustakas},
  {Fassnacht}, {Brada{\v c}}, {Schrabback}, \& {Blandford}}]{Marshall09}
{Marshall}, P.~J., {Hogg}, D.~W., {Moustakas}, L.~A., {et~al.} 2009, \apj, 694,
  924

\bibitem[{{Martizzi} {et~al.}(2012){Martizzi}, {Teyssier}, {Moore}, \&
  {Wentz}}]{Martizzi12}
{Martizzi}, D., {Teyssier}, R., {Moore}, B., \& {Wentz}, T. 2012, \mnras, 422,
  3081

\bibitem[{{McCully} {et~al.}(2014){McCully}, {Jha}, {Foley}, {Bildsten},
  {Fong}, {Kirshner}, {Marion}, {Riess}, \& {Stritzinger}}]{McCully14}
{McCully}, C., {Jha}, S.~W., {Foley}, R.~J., {et~al.} 2014, \nat, 512, 54

\bibitem[{{McKean} {et~al.}(2015){McKean}, {Jackson}, {Vegetti}, {Rybak},
  {Serjeant}, {Koopmans}, {Metcalf}, {Fassnacht}, {Marshall}, \&
  {Pandey-Pommier}}]{McKean15}
{McKean}, J., {Jackson}, N., {Vegetti}, S., {et~al.} 2015, Advancing
  Astrophysics with the Square Kilometre Array (AASKA14), 84

\bibitem[{{Melinder} {et~al.}(2012){Melinder}, {Dahlen}, {Menc{\'{\i}}a
  Trinchant}, {{\"O}stlin}, {Mattila}, {Sollerman}, {Fransson}, {Hayes},
  {Kankare}, \& {Nasoudi-Shoar}}]{Melinder12}
{Melinder}, J., {Dahlen}, T., {Menc{\'{\i}}a Trinchant}, L., {et~al.} 2012,
  \aap, 545, A96

\bibitem[{{More} {et~al.}(2012){More}, {Cabanac}, {More}, {Alard}, {Limousin},
  {Kneib}, {Gavazzi}, \& {Motta}}]{More12}
{More}, A., {Cabanac}, R., {More}, S., {et~al.} 2012, \apj, 749, 38

\bibitem[{{More} {et~al.}(2017){More}, {Suyu}, {Oguri}, {More}, \&
  {Lee}}]{More17}
{More}, A., {Suyu}, S.~H., {Oguri}, M., {More}, S., \& {Lee}, C.-H. 2017,
  \apjl, 835, L25

\bibitem[{{More} {et~al.}(2016){More}, {Verma}, {Marshall}, {More}, {Baeten},
  {Wilcox}, {Macmillan}, {Cornen}, {Kapadia}, {Parrish}, {Snyder}, {Davis},
  {Gavazzi}, {Lintott}, {Simpson}, {Miller}, {Smith}, {Paget}, {Saha},
  {K{\"u}ng}, \& {Collett}}]{More16a}
{More}, A., {Verma}, A., {Marshall}, P.~J., {et~al.} 2016, \mnras, 455, 1191

\bibitem[{{Morgan} {et~al.}(2012){Morgan}, {Hainline}, {Chen}, {Tewes},
  {Kochanek}, {Dai}, {Kozlowski}, {Blackburne}, {Mosquera}, {Chartas},
  {Courbin}, \& {Meylan}}]{Morgan12}
{Morgan}, C.~W., {Hainline}, L.~J., {Chen}, B., {et~al.} 2012, \apj, 756, 52

\bibitem[{{Morozova} {et~al.}(2017){Morozova}, {Piro}, \&
  {Valenti}}]{Morozova17}
{Morozova}, V., {Piro}, A.~L., \& {Valenti}, S. 2017, \apj, 838, 28

\bibitem[{{M{\"o}rtsell} {et~al.}(2005){M{\"o}rtsell}, {Dahle}, \&
  {Hannestad}}]{Mortsell05}
{M{\"o}rtsell}, E., {Dahle}, H., \& {Hannestad}, S. 2005, \apj, 619, 733

\bibitem[{{Mostek} {et~al.}(2012){Mostek}, {Coil}, {Moustakas}, {Salim}, \&
  {Weiner}}]{Mostek12}
{Mostek}, N., {Coil}, A.~L., {Moustakas}, J., {Salim}, S., \& {Weiner}, B.~J.
  2012, \apj, 746, 124

\bibitem[{{Moustakas} {et~al.}(2006){Moustakas}, {Kennicutt}, \&
  {Tremonti}}]{Moustakas06}
{Moustakas}, J., {Kennicutt}, Jr., R.~C., \& {Tremonti}, C.~A. 2006, \apj, 642,
  775

\bibitem[{{Mu{\~n}oz} {et~al.}(1998){Mu{\~n}oz}, {Falco}, {Kochanek},
  {Leh{\'a}r}, {McLeod}, {Impey}, {Rix}, \& {Peng}}]{Munoz98}
{Mu{\~n}oz}, J.~A., {Falco}, E.~E., {Kochanek}, C.~S., {et~al.} 1998, \apss,
  263, 51

\bibitem[{{Muzzin} {et~al.}(2012){Muzzin}, {Labb{\'e}}, {Franx}, {van Dokkum},
  {Holt}, {Szomoru}, {van de Sande}, {Brammer}, {Marchesini}, {Stefanon},
  {Buitrago}, {Caputi}, {Dunlop}, {Fynbo}, {Le F{\'e}vre}, {McCracken}, \&
  {Milvang-Jensen}}]{Muzzin12}
{Muzzin}, A., {Labb{\'e}}, I., {Franx}, M., {et~al.} 2012, \apj, 761, 142

\bibitem[{{Negrello} {et~al.}(2017){Negrello}, {Amber}, {Amvrosiadis}, {Cai},
  {Lapi}, {Gonzalez-Nuevo}, {De Zotti}, {Furlanetto}, {Maddox}, {Allen},
  {Bakx}, {Bussmann}, {Cooray}, {Covone}, {Danese}, {Dannerbauer}, {Fu},
  {Greenslade}, {Gurwell}, {Hopwood}, {Koopmans}, {Napolitano}, {Nayyeri},
  {Omont}, {Petrillo}, {Riechers}, {Serjeant}, {Tortora}, {Valiante}, {Verdoes
  Kleijn}, {Vernardos}, {Wardlow}, {Baes}, {Baker}, {Bourne}, {Clements},
  {Crawford}, {Dye}, {Dunne}, {Eales}, {Ivison}, {Marchetti}, {Micha{\l}owski},
  {Smith}, {Vaccari}, \& {van der Werf}}]{Negrello17}
{Negrello}, M., {Amber}, S., {Amvrosiadis}, A., {et~al.} 2017, \mnras, 465,
  3558

\bibitem[{{Nipoti} {et~al.}(2004){Nipoti}, {Treu}, {Ciotti}, \&
  {Stiavelli}}]{Nipoti04}
{Nipoti}, C., {Treu}, T., {Ciotti}, L., \& {Stiavelli}, M. 2004, \mnras, 355,
  1119

\bibitem[{{Nugent} {et~al.}(2006){Nugent}, {Sullivan}, {Ellis}, {Gal-Yam},
  {Leonard}, {Howell}, {Astier}, {Carlberg}, {Conley}, {Fabbro}, {Fouchez},
  {Neill}, {Pain}, {Perrett}, {Pritchet}, \& {Regnault}}]{Nugent06}
{Nugent}, P., {Sullivan}, M., {Ellis}, R., {et~al.} 2006, \apj, 645, 841

\bibitem[{{Nugent} {et~al.}(2011){Nugent}, {Sullivan}, {Cenko}, {Thomas},
  {Kasen}, {Howell}, {Bersier}, {Bloom}, {Kulkarni}, {Kandrashoff},
  {Filippenko}, {Silverman}, {Marcy}, {Howard}, {Isaacson}, {Maguire},
  {Suzuki}, {Tarlton}, {Pan}, {Bildsten}, {Fulton}, {Parrent}, {Sand},
  {Podsiadlowski}, {Bianco}, {Dilday}, {Graham}, {Lyman}, {James}, {Kasliwal},
  {Law}, {Quimby}, {Hook}, {Walker}, {Mazzali}, {Pian}, {Ofek}, {Gal-Yam}, \&
  {Poznanski}}]{Nugent11}
{Nugent}, P.~E., {Sullivan}, M., {Cenko}, S.~B., {et~al.} 2011, \nat, 480, 344

\bibitem[{{Oguri} \& {Kawano}(2003)}]{Oguri03}
{Oguri}, M., \& {Kawano}, Y. 2003, \mnras, 338, L25

\bibitem[{{Oguri} \& {Marshall}(2010)}]{Oguri10}
{Oguri}, M., \& {Marshall}, P.~J. 2010, \mnras, 405, 2579

\bibitem[{{Olivares E.} {et~al.}(2010){Olivares E.}, {Hamuy}, {Pignata},
  {Maza}, {Bersten}, {Phillips}, {Suntzeff}, {Filippenko}, {Morrel},
  {Kirshner}, \& {Matheson}}]{Olivares10}
{Olivares E.}, F., {Hamuy}, M., {Pignata}, G., {et~al.} 2010, \apj, 715, 833

\bibitem[{{Pan} \& {Loeb}(2013)}]{Pan13}
{Pan}, T., \& {Loeb}, A. 2013, \mnras, 435, L33

\bibitem[{{Pawase} {et~al.}(2014){Pawase}, {Courbin}, {Faure}, {Kokotanekova},
  \& {Meylan}}]{Pawase14}
{Pawase}, R.~S., {Courbin}, F., {Faure}, C., {Kokotanekova}, R., \& {Meylan},
  G. 2014, \mnras, 439, 3392

\bibitem[{{Percival} {et~al.}(2010){Percival}, {Reid}, {Eisenstein}, {Bahcall},
  {Budavari}, {Frieman}, {Fukugita}, {Gunn}, {Ivezi{\'c}}, {Knapp}, {Kron},
  {Loveday}, {Lupton}, {McKay}, {Meiksin}, {Nichol}, {Pope}, {Schlegel},
  {Schneider}, {Spergel}, {Stoughton}, {Strauss}, {Szalay}, {Tegmark},
  {Vogeley}, {Weinberg}, {York}, \& {Zehavi}}]{Percival10}
{Percival}, W.~J., {Reid}, B.~A., {Eisenstein}, D.~J., {et~al.} 2010, \mnras,
  401, 2148

\bibitem[{{Perlmutter} {et~al.}(1999){Perlmutter}, {Aldering}, {Goldhaber},
  {Knop}, {Nugent}, {Castro}, {Deustua}, {Fabbro}, {Goobar}, {Groom}, {Hook},
  {Kim}, {Kim}, {Lee}, {Nunes}, {Pain}, {Pennypacker}, {Quimby}, {Lidman},
  {Ellis}, {Irwin}, {McMahon}, {Ruiz-Lapuente}, {Walton}, {Schaefer}, {Boyle},
  {Filippenko}, {Matheson}, {Fruchter}, {Panagia}, {Newberg}, {Couch}, \&
  {Project}}]{Perlmutter99}
{Perlmutter}, S., {Aldering}, G., {Goldhaber}, G., {et~al.} 1999, \apj, 517,
  565

\bibitem[{{Petrushevska} {et~al.}(2016){Petrushevska}, {Amanullah}, {Goobar},
  {Fabbro}, {Johansson}, {Kjellsson}, {Lidman}, {Paech}, {Richard}, {Dahle},
  {Ferretti}, {Kneib}, {Limousin}, {Nordin}, \& {Stanishev}}]{Petrushevska16}
{Petrushevska}, T., {Amanullah}, R., {Goobar}, A., {et~al.} 2016, \aap, 594,
  A54

\bibitem[{{Petrushevska} {et~al.}(2018){Petrushevska}, {Goobar}, {Lagattuta},
  {Amanullah}, {Hangard}, {Fabbro}, {Lidman}, {Paech}, {Richard}, \&
  {Kneib}}]{Petrushevska18}
{Petrushevska}, T., {Goobar}, A., {Lagattuta}, D.~J., {et~al.} 2018, \aap, 614,
  A103

\bibitem[{{Planck Collaboration} {et~al.}(2015){Planck Collaboration}, {Ade},
  {Aghanim}, {Arnaud}, {Ashdown}, {Aumont}, {Baccigalupi}, {Banday},
  {Barreiro}, {Bartlett}, \& et~al.}]{Planck15}
{Planck Collaboration}, {Ade}, P.~A.~R., {Aghanim}, N., {et~al.} 2015, ArXiv
  e-prints, arXiv:1502.01589

\bibitem[{{Postman} {et~al.}(2012){Postman}, {Coe}, {Ben{\'{\i}}tez},
  {Bradley}, {Broadhurst}, {Donahue}, {Ford}, {Graur}, {Graves}, {Jouvel},
  {Koekemoer}, {Lemze}, {Medezinski}, {Molino}, {Moustakas}, {Ogaz}, {Riess},
  {Rodney}, {Rosati}, {Umetsu}, {Zheng}, {Zitrin}, {Bartelmann}, {Bouwens},
  {Czakon}, {Golwala}, {Host}, {Infante}, {Jha}, {Jimenez-Teja}, {Kelson},
  {Lahav}, {Lazkoz}, {Maoz}, {McCully}, {Melchior}, {Meneghetti}, {Merten},
  {Moustakas}, {Nonino}, {Patel}, {Reg{\"o}s}, {Sayers}, {Seitz}, \& {Van der
  Wel}}]{Postman12}
{Postman}, M., {Coe}, D., {Ben{\'{\i}}tez}, N., {et~al.} 2012, \apjs, 199, 25

\bibitem[{{Poznanski} {et~al.}(2009){Poznanski}, {Butler}, {Filippenko},
  {Ganeshalingam}, {Li}, {Bloom}, {Chornock}, {Foley}, {Nugent}, {Silverman},
  {Cenko}, {Gates}, {Leonard}, {Miller}, {Modjaz}, {Serduke}, {Smith}, {Swift},
  \& {Wong}}]{Poznanski09}
{Poznanski}, D., {Butler}, N., {Filippenko}, A.~V., {et~al.} 2009, \apj, 694,
  1067

\bibitem[{{Pritchard} {et~al.}(2014){Pritchard}, {Roming}, {Brown}, {Bayless},
  \& {Frey}}]{Pritchard14}
{Pritchard}, T.~A., {Roming}, P.~W.~A., {Brown}, P.~J., {Bayless}, A.~J., \&
  {Frey}, L.~H. 2014, \apj, 787, 157

\bibitem[{{Pugh} {et~al.}(2004){Pugh}, {Park}, \& {Li}}]{Pugh04}
{Pugh}, H., {Park}, S., \& {Li}, W. 2004, \iaucirc, 8425

\bibitem[{{Quider} {et~al.}(2009){Quider}, {Pettini}, {Shapley}, \&
  {Steidel}}]{Quider09}
{Quider}, A.~M., {Pettini}, M., {Shapley}, A.~E., \& {Steidel}, C.~C. 2009,
  \mnras, 398, 1263

\bibitem[{{Quimby} {et~al.}(2014){Quimby}, {Oguri}, {More}, {More}, {Moriya},
  {Werner}, {Tanaka}, {Folatelli}, {Bersten}, {Maeda}, \& {Nomoto}}]{Quimby14}
{Quimby}, R.~M., {Oguri}, M., {More}, A., {et~al.} 2014, Science, 344, 396

\bibitem[{{Richardson} {et~al.}(2002){Richardson}, {Branch}, {Casebeer},
  {Millard}, {Thomas}, \& {Baron}}]{Richardson02}
{Richardson}, D., {Branch}, D., {Casebeer}, D., {et~al.} 2002, \aj, 123, 745

\bibitem[{{Richardson} {et~al.}(2014){Richardson}, {Jenkins}, {Wright}, \&
  {Maddox}}]{Richardson14}
{Richardson}, D., {Jenkins}, III, R.~L., {Wright}, J., \& {Maddox}, L. 2014,
  \aj, 147, 118

\bibitem[{{Richmond} {et~al.}(1998){Richmond}, {Filippenko}, \&
  {Galisky}}]{Richmond98}
{Richmond}, M.~W., {Filippenko}, A.~V., \& {Galisky}, J. 1998, \pasp, 110, 553

\bibitem[{{Riess} {et~al.}(1998){Riess}, {Filippenko}, {Challis},
  {Clocchiatti}, {Diercks}, {Garnavich}, {Gilliland}, {Hogan}, {Jha},
  {Kirshner}, {Leibundgut}, {Phillips}, {Reiss}, {Schmidt}, {Schommer},
  {Smith}, {Spyromilio}, {Stubbs}, {Suntzeff}, \& {Tonry}}]{Riess98}
{Riess}, A.~G., {Filippenko}, A.~V., {Challis}, P., {et~al.} 1998, \aj, 116,
  1009

\bibitem[{{Rodney} {et~al.}(2015){Rodney}, {Patel}, {Scolnic}, {Foley},
  {Molino}, {Brammer}, {Jauzac}, {Brada{\v c}}, {Broadhurst}, {Coe}, {Diego},
  {Graur}, {Hjorth}, {Hoag}, {Jha}, {Johnson}, {Kelly}, {Lam}, {McCully},
  {Medezinski}, {Meneghetti}, {Merten}, {Richard}, {Riess}, {Sharon},
  {Strolger}, {Treu}, {Wang}, {Williams}, \& {Zitrin}}]{Rodney15}
{Rodney}, S.~A., {Patel}, B., {Scolnic}, D., {et~al.} 2015, \apj, 811, 70

\bibitem[{{Romano-D{\'{\i}}az} {et~al.}(2008){Romano-D{\'{\i}}az}, {Shlosman},
  {Hoffman}, \& {Heller}}]{Romano-Diaz08}
{Romano-D{\'{\i}}az}, E., {Shlosman}, I., {Hoffman}, Y., \& {Heller}, C. 2008,
  \apjl, 685, L105

\bibitem[{{Rubin} \& {Hayden}(2016)}]{Rubin16}
{Rubin}, D., \& {Hayden}, B. 2016, \apjl, 833, L30

\bibitem[{{Saha}(2000)}]{Saha00}
{Saha}, P. 2000, \aj, 120, 1654

\bibitem[{{Sako} {et~al.}(2014){Sako}, {Bassett}, {Becker}, {Brown},
  {Campbell}, {Cane}, {Cinabro}, {D'Andrea}, {Dawson}, {DeJongh}, {Depoy},
  {Dilday}, {Doi}, {Filippenko}, {Fischer}, {Foley}, {Frieman}, {Galbany},
  {Garnavich}, {Goobar}, {Gupta}, {Hill}, {Hayden}, {Hlozek}, {Holtzman},
  {Hopp}, {Jha}, {Kessler}, {Kollatschny}, {Leloudas}, {Marriner}, {Marshall},
  {Miquel}, {Morokuma}, {Mosher}, {Nichol}, {Nordin}, {Olmstead}, {Ostman},
  {Prieto}, {Richmond}, {Romani}, {Sollerman}, {Stritzinger}, {Schneider},
  {Smith}, {Wheeler}, {Yasuda}, \& {Zheng}}]{Sako14}
{Sako}, M., {Bassett}, B., {Becker}, A.~C., {et~al.} 2014, ArXiv e-prints,
  arXiv:1401.3317

\bibitem[{{Salpeter}(1955)}]{Salpeter55}
{Salpeter}, E.~E. 1955, \apj, 121, 161

\bibitem[{{Sanders} {et~al.}(2015){Sanders}, {Soderberg}, {Gezari},
  {Betancourt}, {Chornock}, {Berger}, {Foley}, {Challis}, {Drout}, {Kirshner},
  {Lunnan}, {Marion}, {Margutti}, {McKinnon}, {Milisavljevic}, {Narayan},
  {Rest}, {Kankare}, {Mattila}, {Smartt}, {Huber}, {Burgett}, {Draper},
  {Hodapp}, {Kaiser}, {Kudritzki}, {Magnier}, {Metcalfe}, {Morgan}, {Price},
  {Tonry}, {Wainscoat}, \& {Waters}}]{Sanders15}
{Sanders}, N.~E., {Soderberg}, A.~M., {Gezari}, S., {et~al.} 2015, \apj, 799,
  208

\bibitem[{{Schmidt} {et~al.}(1998){Schmidt}, {Suntzeff}, {Phillips},
  {Schommer}, {Clocchiatti}, {Kirshner}, {Garnavich}, {Challis}, {Leibundgut},
  {Spyromilio}, {Riess}, {Filippenko}, {Hamuy}, {Smith}, {Hogan}, {Stubbs},
  {Diercks}, {Reiss}, {Gilliland}, {Tonry}, {Maza}, {Dressler}, {Walsh}, \&
  {Ciardullo}}]{Schmidt98}
{Schmidt}, B.~P., {Suntzeff}, N.~B., {Phillips}, M.~M., {et~al.} 1998, \apj,
  507, 46

\bibitem[{{Schneider} \& {Sluse}(2013)}]{Schneider13}
{Schneider}, P., \& {Sluse}, D. 2013, \aap, 559, A37

\bibitem[{{Schneider} \& {Sluse}(2014)}]{Schneider14}
---. 2014, \aap, 564, A103

\bibitem[{{Shu} {et~al.}(2016{\natexlab{a}}){Shu}, {Bolton}, {Moustakas},
  {Stern}, {Dey}, {Brownstein}, {Burles}, \& {Spinrad}}]{Shu16a}
{Shu}, Y., {Bolton}, A.~S., {Moustakas}, L.~A., {et~al.} 2016{\natexlab{a}},
  \apj, 820, 43

\bibitem[{{Shu} {et~al.}(2015){Shu}, {Bolton}, {Brownstein}, {Montero-Dorta},
  {Koopmans}, {Treu}, {Gavazzi}, {Auger}, {Czoske}, {Marshall}, \&
  {Moustakas}}]{Shu15}
{Shu}, Y., {Bolton}, A.~S., {Brownstein}, J.~R., {et~al.} 2015, \apj, 803, 71

\bibitem[{{Shu} {et~al.}(2016{\natexlab{b}}){Shu}, {Bolton}, {Kochanek},
  {Oguri}, {P{\'e}rez-Fournon}, {Zheng}, {Mao}, {Montero-Dorta}, {Brownstein},
  {Marques-Chaves}, \& {M{\'e}nard}}]{BELLSIII}
{Shu}, Y., {Bolton}, A.~S., {Kochanek}, C.~S., {et~al.} 2016{\natexlab{b}},
  \apj, 824, 86

\bibitem[{{Shu} {et~al.}(2016{\natexlab{c}}){Shu}, {Bolton}, {Mao}, {Kochanek},
  {P{\'e}rez-Fournon}, {Oguri}, {Montero-Dorta}, {Cornachione},
  {Marques-Chaves}, {Zheng}, {Brownstein}, \& {M{\'e}nard}}]{BELLS_IV}
{Shu}, Y., {Bolton}, A.~S., {Mao}, S., {et~al.} 2016{\natexlab{c}}, \apj, 833,
  264

\bibitem[{{Shu} {et~al.}(2017){Shu}, {Brownstein}, {Bolton}, {Koopmans},
  {Treu}, {Montero-Dorta}, {Auger}, {Czoske}, {Gavazzi}, {Marshall}, \&
  {Moustakas}}]{SLACSXIII}
{Shu}, Y., {Brownstein}, J.~R., {Bolton}, A.~S., {et~al.} 2017, \apj, 851, 48

\bibitem[{{Smith} {et~al.}(2012){Smith}, {Nichol}, {Dilday}, {Marriner},
  {Kessler}, {Bassett}, {Cinabro}, {Frieman}, {Garnavich}, {Jha}, {Lampeitl},
  {Sako}, {Schneider}, \& {Sollerman}}]{Smith12}
{Smith}, M., {Nichol}, R.~C., {Dilday}, B., {et~al.} 2012, \apj, 755, 61

\bibitem[{{Sobral} {et~al.}(2012){Sobral}, {Best}, {Matsuda}, {Smail}, {Geach},
  \& {Cirasuolo}}]{Sobral12}
{Sobral}, D., {Best}, P.~N., {Matsuda}, Y., {et~al.} 2012, \mnras, 420, 1926

\bibitem[{{Soderberg} {et~al.}(2008){Soderberg}, {Berger}, {Page}, {Schady},
  {Parrent}, {Pooley}, {Wang}, {Ofek}, {Cucchiara}, {Rau}, {Waxman}, {Simon},
  {Bock}, {Milne}, {Page}, {Barentine}, {Barthelmy}, {Beardmore}, {Bietenholz},
  {Brown}, {Burrows}, {Burrows}, {Byrngelson}, {Cenko}, {Chandra}, {Cummings},
  {Fox}, {Gal-Yam}, {Gehrels}, {Immler}, {Kasliwal}, {Kong}, {Krimm},
  {Kulkarni}, {Maccarone}, {M{\'e}sz{\'a}ros}, {Nakar}, {O'Brien}, {Overzier},
  {de Pasquale}, {Racusin}, {Rea}, \& {York}}]{Soderberg08}
{Soderberg}, A.~M., {Berger}, E., {Page}, K.~L., {et~al.} 2008, \nat, 453, 469

\bibitem[{{Sonnenfeld} {et~al.}(2013){Sonnenfeld}, {Gavazzi}, {Suyu}, {Treu},
  \& {Marshall}}]{Sonnenfeld13}
{Sonnenfeld}, A., {Gavazzi}, R., {Suyu}, S.~H., {Treu}, T., \& {Marshall},
  P.~J. 2013, \apj, 777, 97

\bibitem[{{Sonnenfeld} {et~al.}(2012){Sonnenfeld}, {Treu}, {Gavazzi},
  {Marshall}, {Auger}, {Suyu}, {Koopmans}, \& {Bolton}}]{Sonnenfeld12}
{Sonnenfeld}, A., {Treu}, T., {Gavazzi}, R., {et~al.} 2012, \apj, 752, 163

\bibitem[{{Sonnenfeld} {et~al.}(2017){Sonnenfeld}, {Chan}, {Shu}, {More},
  {Oguri}, {Suyu}, {Wong}, {Lee}, {Coupon}, {Yonehara}, {Bolton}, {Jaelani},
  {Tanaka}, {Miyazaki}, \& {Komiyama}}]{Sonnenfeld17}
{Sonnenfeld}, A., {Chan}, J.~H.~H., {Shu}, Y., {et~al.} 2017, ArXiv e-prints,
  arXiv:1704.01585

\bibitem[{{Spilker} {et~al.}(2016){Spilker}, {Marrone}, {Aravena},
  {B{\'e}thermin}, {Bothwell}, {Carlstrom}, {Chapman}, {Crawford}, {de Breuck},
  {Fassnacht}, {Gonzalez}, {Greve}, {Hezaveh}, {Litke}, {Ma}, {Malkan},
  {Rotermund}, {Strandet}, {Vieira}, {Weiss}, \& {Welikala}}]{Spilker16}
{Spilker}, J.~S., {Marrone}, D.~P., {Aravena}, M., {et~al.} 2016, \apj, 826,
  112

\bibitem[{{Spiniello} {et~al.}(2014){Spiniello}, {Trager}, {Koopmans}, \&
  {Conroy}}]{Spiniello14}
{Spiniello}, C., {Trager}, S., {Koopmans}, L.~V.~E., \& {Conroy}, C. 2014,
  \mnras, 438, 1483

\bibitem[{{Spiniello} {et~al.}(2012){Spiniello}, {Trager}, {Koopmans}, \&
  {Chen}}]{Spiniello12}
{Spiniello}, C., {Trager}, S.~C., {Koopmans}, L.~V.~E., \& {Chen}, Y.~P. 2012,
  \apjl, 753, L32

\bibitem[{{Stark} {et~al.}(2013){Stark}, {Auger}, {Belokurov}, {Jones},
  {Robertson}, {Ellis}, {Sand}, {Moiseev}, {Eagle}, \& {Myers}}]{Stark13}
{Stark}, D.~P., {Auger}, M., {Belokurov}, V., {et~al.} 2013, \mnras, 436, 1040

\bibitem[{{Stark} {et~al.}(2015){Stark}, {Walth}, {Charlot}, {Cl{\'e}ment},
  {Feltre}, {Gutkin}, {Richard}, {Mainali}, {Robertson}, {Siana}, {Tang}, \&
  {Schenker}}]{Stark15}
{Stark}, D.~P., {Walth}, G., {Charlot}, S., {et~al.} 2015, \mnras, 454, 1393

\bibitem[{{Strader} {et~al.}(2011){Strader}, {Caldwell}, \& {Seth}}]{Strader11}
{Strader}, J., {Caldwell}, N., \& {Seth}, A.~C. 2011, \aj, 142, 8

\bibitem[{{Strolger} {et~al.}(2015){Strolger}, {Dahlen}, {Rodney}, {Graur},
  {Riess}, {McCully}, {Ravindranath}, {Mobasher}, \& {Shahady}}]{Strolger15}
{Strolger}, L.-G., {Dahlen}, T., {Rodney}, S.~A., {et~al.} 2015, \apj, 813, 93

\bibitem[{{Sullivan} {et~al.}(2006){Sullivan}, {Le Borgne}, {Pritchet},
  {Hodsman}, {Neill}, {Howell}, {Carlberg}, {Astier}, {Aubourg}, {Balam},
  {Basa}, {Conley}, {Fabbro}, {Fouchez}, {Guy}, {Hook}, {Pain},
  {Palanque-Delabrouille}, {Perrett}, {Regnault}, {Rich}, {Taillet}, {Baumont},
  {Bronder}, {Ellis}, {Filiol}, {Lusset}, {Perlmutter}, {Ripoche}, \&
  {Tao}}]{Sullivan06}
{Sullivan}, M., {Le Borgne}, D., {Pritchet}, C.~J., {et~al.} 2006, \apj, 648,
  868

\bibitem[{{Suyu} {et~al.}(2010){Suyu}, {Marshall}, {Auger}, {Hilbert},
  {Blandford}, {Koopmans}, {Fassnacht}, \& {Treu}}]{Suyu10a}
{Suyu}, S.~H., {Marshall}, P.~J., {Auger}, M.~W., {et~al.} 2010, \apj, 711, 201

\bibitem[{{Suyu} {et~al.}(2013){Suyu}, {Auger}, {Hilbert}, {Marshall}, {Tewes},
  {Treu}, {Fassnacht}, {Koopmans}, {Sluse}, {Blandford}, {Courbin}, \&
  {Meylan}}]{Suyu13}
{Suyu}, S.~H., {Auger}, M.~W., {Hilbert}, S., {et~al.} 2013, \apj, 766, 70

\bibitem[{{Suyu} {et~al.}(2014){Suyu}, {Treu}, {Hilbert}, {Sonnenfeld},
  {Auger}, {Blandford}, {Collett}, {Courbin}, {Fassnacht}, {Koopmans},
  {Marshall}, {Meylan}, {Spiniello}, \& {Tewes}}]{Suyu14}
{Suyu}, S.~H., {Treu}, T., {Hilbert}, S., {et~al.} 2014, \apjl, 788, L35

\bibitem[{{Taddia} {et~al.}(2018){Taddia}, {Stritzinger}, {Bersten}, {Baron},
  {Burns}, {Contreras}, {Holmbo}, {Hsiao}, {Morrell}, {Phillips}, {Sollerman},
  \& {Suntzeff}}]{Taddia18}
{Taddia}, F., {Stritzinger}, M.~D., {Bersten}, M., {et~al.} 2018, \aap, 609,
  A136

\bibitem[{{Talia} {et~al.}(2015){Talia}, {Cimatti}, {Pozzetti}, {Rodighiero},
  {Gruppioni}, {Pozzi}, {Daddi}, {Maraston}, {Mignoli}, \& {Kurk}}]{Talia15}
{Talia}, M., {Cimatti}, A., {Pozzetti}, L., {et~al.} 2015, \aap, 582, A80

\bibitem[{{Tammann} {et~al.}(1994){Tammann}, {Loeffler}, \&
  {Schroeder}}]{Tammann94}
{Tammann}, G.~A., {Loeffler}, W., \& {Schroeder}, A. 1994, \apjs, 92, 487

\bibitem[{{Tewes} {et~al.}(2013){Tewes}, {Courbin}, \& {Meylan}}]{Tewes13}
{Tewes}, M., {Courbin}, F., \& {Meylan}, G. 2013, \aap, 553, A120

\bibitem[{{Tie} \& {Kochanek}(2018)}]{Tie18}
{Tie}, S.~S., \& {Kochanek}, C.~S. 2018, \mnras, 473, 80

\bibitem[{{Tornambe'} {et~al.}(2017){Tornambe'}, {Piersanti}, {Raimondo}, \&
  {Delgrande}}]{Tornambe17}
{Tornambe'}, A., {Piersanti}, L., {Raimondo}, G., \& {Delgrande}, R. 2017,
  ArXiv e-prints, arXiv:1712.09810

\bibitem[{{Tortora} {et~al.}(2013){Tortora}, {Romanowsky}, \&
  {Napolitano}}]{Tortora13}
{Tortora}, C., {Romanowsky}, A.~J., \& {Napolitano}, N.~R. 2013, \apj, 765, 8

\bibitem[{{Treu} {et~al.}(2010){Treu}, {Auger}, {Koopmans}, {Gavazzi},
  {Marshall}, \& {Bolton}}]{Treu10a}
{Treu}, T., {Auger}, M.~W., {Koopmans}, L.~V.~E., {et~al.} 2010, \apj, 709,
  1195

\bibitem[{{Treu} {et~al.}(2011){Treu}, {Dutton}, {Auger}, {Marshall}, {Bolton},
  {Brewer}, {Koo}, \& {Koopmans}}]{SWELLSI}
{Treu}, T., {Dutton}, A.~A., {Auger}, M.~W., {et~al.} 2011, \mnras, 417, 1601

\bibitem[{{Treu} \& {Koopmans}(2002)}]{Treu02}
{Treu}, T., \& {Koopmans}, L.~V.~E. 2002, \mnras, 337, L6

\bibitem[{{Tsujimoto} {et~al.}(1997){Tsujimoto}, {Yoshii}, {Nomoto},
  {Matteucci}, {Thielemann}, \& {Hashimoto}}]{Tsujimoto97}
{Tsujimoto}, T., {Yoshii}, Y., {Nomoto}, K., {et~al.} 1997, \apj, 483, 228

\bibitem[{{van Dokkum} \& {Conroy}(2010)}]{vanDokkum10}
{van Dokkum}, P.~G., \& {Conroy}, C. 2010, \nat, 468, 940

\bibitem[{{Velliscig} {et~al.}(2014){Velliscig}, {van Daalen}, {Schaye},
  {McCarthy}, {Cacciato}, {Le Brun}, \& {Dalla Vecchia}}]{Velliscig14}
{Velliscig}, M., {van Daalen}, M.~P., {Schaye}, J., {et~al.} 2014, \mnras, 442,
  2641

\bibitem[{{Vieira} {et~al.}(2013){Vieira}, {Marrone}, {Chapman}, {De Breuck},
  {Hezaveh}, {Wei{$\beta$}}, {Aguirre}, {Aird}, {Aravena}, {Ashby}, {Bayliss},
  {Benson}, {Biggs}, {Bleem}, {Bock}, {Bothwell}, {Bradford}, {Brodwin},
  {Carlstrom}, {Chang}, {Crawford}, {Crites}, {de Haan}, {Dobbs}, {Fomalont},
  {Fassnacht}, {George}, {Gladders}, {Gonzalez}, {Greve}, {Gullberg},
  {Halverson}, {High}, {Holder}, {Holzapfel}, {Hoover}, {Hrubes}, {Hunter},
  {Keisler}, {Lee}, {Leitch}, {Lueker}, {Luong-van}, {Malkan}, {McIntyre},
  {McMahon}, {Mehl}, {Menten}, {Meyer}, {Mocanu}, {Murphy}, {Natoli}, {Padin},
  {Plagge}, {Reichardt}, {Rest}, {Ruel}, {Ruhl}, {Sharon}, {Schaffer}, {Shaw},
  {Shirokoff}, {Spilker}, {Stalder}, {Staniszewski}, {Stark}, {Story},
  {Vanderlinde}, {Welikala}, \& {Williamson}}]{Vieira13}
{Vieira}, J.~D., {Marrone}, D.~P., {Chapman}, S.~C., {et~al.} 2013, \nat, 495,
  344

\bibitem[{{Walsh} {et~al.}(1979){Walsh}, {Carswell}, \& {Weymann}}]{Walsh79}
{Walsh}, D., {Carswell}, R.~F., \& {Weymann}, R.~J. 1979, \nat, 279, 381

\bibitem[{{Wang} {et~al.}(2013){Wang}, {Wang}, {Filippenko}, {Zhang}, \&
  {Zhao}}]{Wang13}
{Wang}, X., {Wang}, L., {Filippenko}, A.~V., {Zhang}, T., \& {Zhao}, X. 2013,
  Science, 340, 170

\bibitem[{{Wang} {et~al.}(2017){Wang}, {Zhao}, {Chuang}, {Ross}, {Percival},
  {Gil-Mar{\'{\i}}n}, {Cuesta}, {Kitaura}, {Rodriguez-Torres}, {Brownstein},
  {Eisenstein}, {Ho}, {Kneib}, {Olmstead}, {Prada}, {Rossi}, {S{\'a}nchez},
  {Salazar-Albornoz}, {Thomas}, {Tinker}, {Tojeiro}, {Vargas-Maga{\~n}a}, \&
  {Zhu}}]{Wang17}
{Wang}, Y., {Zhao}, G.-B., {Chuang}, C.-H., {et~al.} 2017, \mnras, 469, 3762

\bibitem[{{Witt} {et~al.}(2000){Witt}, {Mao}, \& {Keeton}}]{Witt00}
{Witt}, H.~J., {Mao}, S., \& {Keeton}, C.~R. 2000, \apj, 544, 98

\bibitem[{{Witt} {et~al.}(1995){Witt}, {Mao}, \& {Schechter}}]{Witt95}
{Witt}, H.~J., {Mao}, S., \& {Schechter}, P.~L. 1995, \apj, 443, 18

\bibitem[{{Wong} {et~al.}(2017){Wong}, {Suyu}, {Auger}, {Bonvin}, {Courbin},
  {Fassnacht}, {Halkola}, {Rusu}, {Sluse}, {Sonnenfeld}, {Treu}, {Collett},
  {Hilbert}, {Koopmans}, {Marshall}, \& {Rumbaugh}}]{Wong17}
{Wong}, K.~C., {Suyu}, S.~H., {Auger}, M.~W., {et~al.} 2017, \mnras, 465, 4895

\bibitem[{{Wucknitz}(2002)}]{Wucknitz02}
{Wucknitz}, O. 2002, \mnras, 332, 951

\bibitem[{{Wyithe} \& {Turner}(2002)}]{Wyithe02}
{Wyithe}, J.~S.~B., \& {Turner}, E.~L. 2002, \apj, 575, 650

\bibitem[{{Yahalomi} {et~al.}(2017){Yahalomi}, {Schechter}, \&
  {Wambsganss}}]{Yahalomi17}
{Yahalomi}, D.~A., {Schechter}, P.~L., \& {Wambsganss}, J. 2017, ArXiv
  e-prints, arXiv:1711.07919

\bibitem[{{Zhao} {et~al.}(2017){Zhao}, {Wang}, {Saito}, {Wang}, {Ross},
  {Beutler}, {Grieb}, {Chuang}, {Kitaura}, {Rodriguez-Torres}, {Percival},
  {Brownstein}, {Cuesta}, {Eisenstein}, {Gil-Mar{\'{\i}}n}, {Kneib}, {Nichol},
  {Olmstead}, {Prada}, {Rossi}, {Salazar-Albornoz}, {Samushia}, {S{\'a}nchez},
  {Thomas}, {Tinker}, {Tojeiro}, {Weinberg}, \& {Zhu}}]{Zhao17}
{Zhao}, G.-B., {Wang}, Y., {Saito}, S., {et~al.} 2017, \mnras, 466, 762

\end{thebibliography}

\end{document}